\documentclass[aps,pre,amsmath]{revtex4}
\pdfoutput=1
\usepackage{graphicx}
\usepackage{color}
\usepackage{bm}
\usepackage{epsfig}
\usepackage{latexsym}
\usepackage{pifont}
\usepackage{float}
\usepackage[utf8]{inputenc}
\usepackage{siunitx}
\usepackage{textgreek}
\usepackage{svg}
\begin{document}
\newcommand{\be}{\begin{equation}}
\newcommand{\ee}{\end{equation}}
\newcommand{\bea}{\begin{eqnarray}}
\newcommand{\eea}{\end{eqnarray}}
\newcommand{\nn}{\nonumber}

\title{A simple cognitive model explains movement decisions during schooling in zebrafish}

\author{Lital Oscar$^1$, Liang Li$^{2,3,4}$, Dan Gorbonos$^{2,3,4}$, Iain D. Couzin$^{2,3,4}$ and Nir S. Gov$^1$\footnote{nir.gov@weizmann.ac.il}}
\affiliation{$^1$Department of Chemical and Biological Physics, Weizmann Institute of Science, Rehovot 7610001, Israel \\ $^2$Department of Collective Behaviour, Max Planck Institute of Animal Behavior, 78464 Konstanz, Germany\\ $^3$Centre for the Advanced Study of Collective
Behaviour, University of Konstanz, 78464 Konstanz, Germany\\ $^4$Department of Biology, University of Konstanz, 78464 Konstanz, Germany}

\begin{abstract}
While moving, animals must frequently make decisions about their future travel direction, whether they are alone or in a group. Here we investigate this process for zebrafish ({\it Danio rerio}), which naturally move in cohesive groups. Employing state-of-the-art virtual reality, we study how real fish follow one or several moving, virtual conspecifics. These data are used to inform, and test, a model of social response that includes a process of explicit decision-making, whereby the fish can decide which of the virtual conspecifics to follow, or to follow some average direction. This approach is in contrast with previous models where the direction of motion was based on  a continuous computation, such as directional averaging. Building upon a simplified version of this model [Sridhar et al., 2021], which has been shown to exhibit a spontaneous symmetry-breaking transition
from moving along a “compromise” (average) direction, to deciding on following one of the virtual fish. This previously published simplified version was limited to a one-dimensional projection of the fish motion, while here we present a model that describes the motion of the real fish as it swims freely in two-dimensions. Here, we extend our proposed Ising-like model, which inherently describes a spontaneous symmetry-breaking transition from moving along a ``compromise'' (average) direction, to deciding on following one of the virtual fish. Motivated by experimental observations, the swim speed of the fish in this model uses a burst-and-coast swimming pattern, with the burst frequency being dependent on the distance of the fish from the followed conspecific(s). We demonstrate that this model is able to explain the observed spatial distribution of the real fish behind the virtual conspecifics in the experiments, as a function of their average speed and number. In particular, the model naturally explains the observed critical bifurcations for a freely swimming fish, which appear in the spatial distributions whenever the fish makes a decision to follow only one of the virtual conspecifics, instead of following them as an averaged group. This model can provide the foundation for modeling a cohesive shoal of swimming fish, while explicitly describing their directional decision-making process at the individual level. 
\end{abstract}

\maketitle

\section{Introduction}\label{Introduction}
Understanding how animals move together in groups is a long-standing puzzle \cite{Sumpter2010Collective,n2}. Current theoretical models of this process involve the description of individual agents that move according to simple rules, which embody the social influence exerted by their conspecifics, and external influences such as physical barriers and predators \citep{n2,zienkiewicz2018data,heras2019deep,2020noise,escobedo2020data,2013collective,lei2020}. Within these models, the influences of the neighbors in the group, representing the sensory information, are integrated, and the resulting direction of motion of each agent is calculated. However, during this calculation, the individuals in such model frameworks do not explicitly perform a decision-making process. In other words, the direction of motion of each agent is given by some unique and continuous function of the positions and velocities of the detected (usually neighboring) conspecifics \cite{n8,couzin2002collective,katz2011inferring,laan2017signatures,zienkiewicz2018data,heras2019deep,wang2022impact}. Given two identical conspecifics, the agent in these modelas has no mechanism to spontaneously decide to follow only one of them (at a given time), in contradiction to the observations \cite{sridhar2021geometry}. 

We propose here a theoretical model for how animals respond to social cues, that includes an explicit decision-making process \cite{sridhar2021geometry}. Our current work is based on our theoretical model for how animal groups, and individual animals, make decisions regarding their desired direction of motion while moving towards and choosing between different targets \cite{n5,sridhar2021geometry}. This assumption is supported by neurobiological studies conducted in a range of animals \cite{sarel2017vectorial,hoydal2019object}.
Within this model, the animal's brain is assumed to have neuronal representations of the vectors aimed at the different targets. The neurons (or neuronal groups) that represent each target are treated as ``spins'', in the terminology of statistical physics, with each spin representing either an ``on'' (1) or ``off'' (0) state. The spin state corresponds to the firing state of the neuronal group that it represents. This spin model is a simplification of (and can be mapped to \cite{sridhar2021geometry}) widely used models of neuronal representation of spatial targets \cite{zhang1996representation,york2009recurrent}. The interactions between the spins drives a geometry-based phase transition (from averaging among options to deciding on one), which manifest as bifurcations in the resulting trajectories. This model thus predicts that the animal's direction of travel would tend to be in a compromise (average) direction between the different targets, when they are far, and the relative angle between them is small. However, above a critical relative angle, such as when the animal approaches the targets, the trajectories bifurcate and point toward one target, in the case of only 2 options, and a subset of targets for a larger number of options. This prediction was verified for several types of insects (desert locusts and fruit flies), moving toward stationary targets in a virtual-reality arena \cite{sridhar2021geometry}. 

The model was also employed to consider how zebrafish (\textit{Danio rerio}) respond to conspecifics. Again using immersive virtual reality \cite{sridhar2021geometry}, allowing virtual fish to be projected in specific configurations while on the move, it was also found that individuals would average when the angle subtended by virtual fish was below a critical value, and a decision to follow one among the remaining options occurred when this angle exceeded a critical value. In the comparison with the theoretical model in this previous work \cite{sridhar2021geometry}, it was assumed that the real fish (RF) can only move along one dimension, perpendicular to the direction of motion of the virtual fish (VF) (Fig.~\ref{nemo}). This way, the RF is assumed to be moving at a fixed distance behind the line of VF, and only its lateral motion sideways was modeled. Note that in the experiments, the RF also maintains a very similar depth to the VF that it is chasing \cite{sridhar2021geometry}, thus schools of real juvenile zebrafish tend to be quasi-planar. Despite the considerable simplification in the theoretical model, the experimental and simulated distributions of the RF, when projected to one dimension, were in very good agreement for both two and three VF \cite{sridhar2021geometry}. 
The essence of the simplification obtained by the one-dimensional projection of the fish motion is that it eliminates the need to calculate the speed of the fish, as it responds to the VF. 

In the experiments, however, the RF moves freely in (predominantly) two dimensions (due to the nature of fish to swim at the same depth as each other) when schooling with the VF, and its distance to the VF constantly changes during the chase. Here we take this freedom of movement into account and develop a theoretical framework that implements the spin model for directional decision-making, together with a calculation of the fish speed, as the RF interacts freely with an arbitrary number and spatial configurations of moving conspecifics (which can be thought of as mobile targets). This model allows us for the first time to compare the spin model for directional decision-making to the full two-dimensional experimental trajectories of the RF \cite{sridhar2021geometry}. We emphasize that our model does not include any explicit alignment interaction of the Vicsek-type \cite{n2,argun2021vicsek}. Moreover, the main theoretical novelty of our work is the spin-based directional decision making model, with its spontaneous symmetry-breaking property, as opposed to previous models using vectorial averaging and simple distance-based interaction rules. The calculation of the speed, using burst-and-coast dynamics, is more standard \cite{harpaz2017discrete,sbragaglia2022evolutionary}.

\section{The Fish Model}

Motivated by our experimental observations on zebrafish (\textit{Danio rerio})  \cite{sridhar2021geometry}, we will limit the motion of the real fish (RF) in our model to a two-dimensional plane on which fish interact (i.e. they are assumed to be swimming at the same depth in the water column) (Fig.~\ref{nemo}). In our experiments the VF move with the burst-and-coast (or, burst-and-glide) motions as do real fish (Fig.~\ref{oneVF}A). In these experiments the virtual fish move in straight trajectories, and make U-turns when they near the edge of the experimental arena. Our model allows us to calculate the full two dimensional motion of the RF, when it interacts with several VF (which move as in the experiments), as illustrated in Fig.~\ref{nemo}. We clearly denote in all the figures the experimental data versus the theoretical results, to allow a direct visual comparison to be made.

\subsection{Direction of motion: Spin dynamics}

\begin{figure}[H]
\centering
{\includegraphics[scale=0.40]{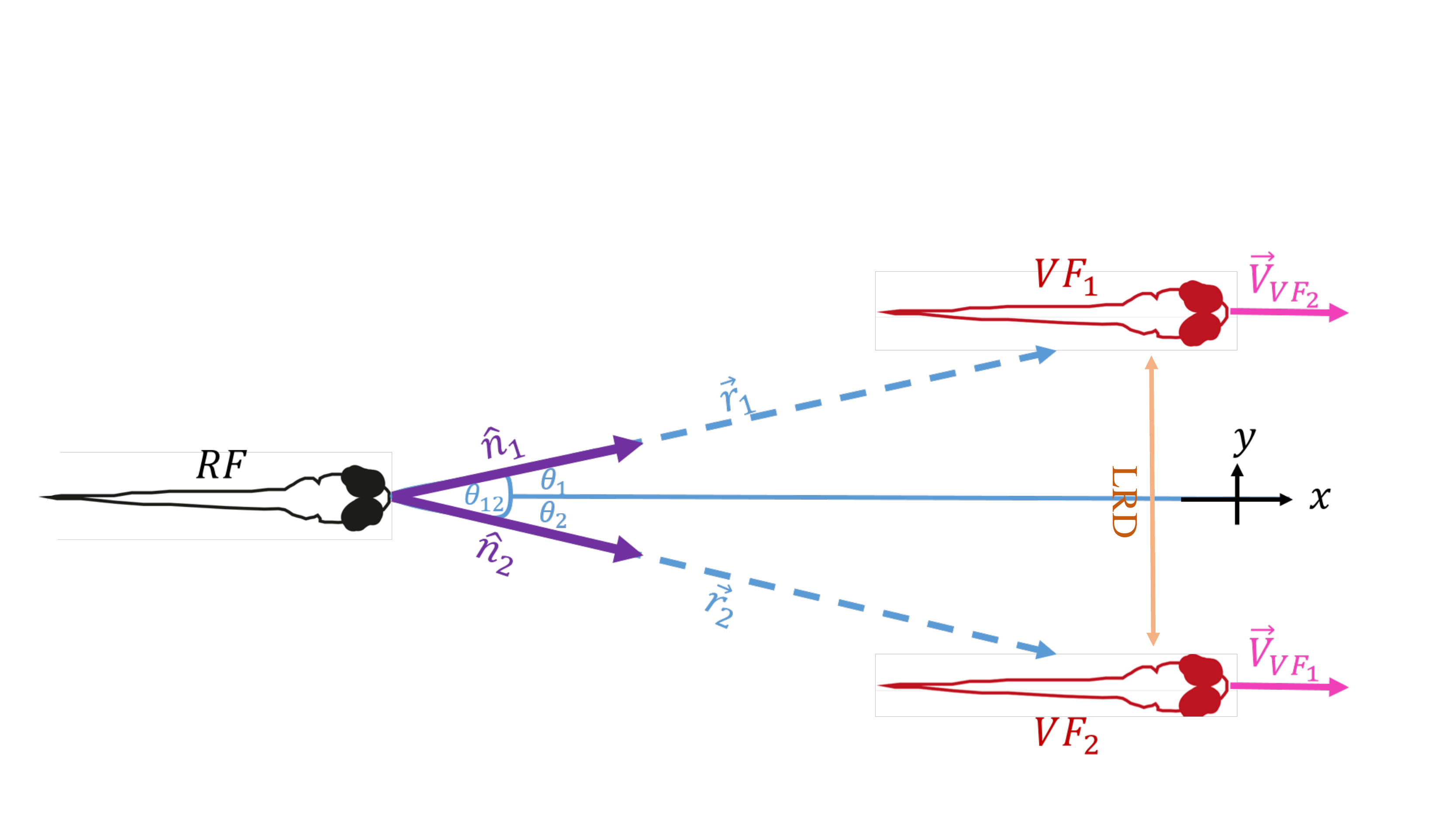}}
\caption{Illustration of a 2D system with 1RF (black) following 2VF (red): $\theta_{12}$ is the instantaneous relative angle between the two targets, $\theta_i$ is the angle from the RF to $VF_i$. 
Note that we use the same terminology to describe our model as employed in our experiments; thus, the RF in the model represents the focal individual whose cognitive process is being simulated whereas the VF represent the VF used in the virtual-reality experiments, moving in the same direction as each other and at a fixed distance apart. The instantaneous direction vectors to the targets are $\hat{n}_1$ and $\hat{n}_2$, with the corresponding distances $r_1$,$r_2$. The lateral distance (LRD) denotes the distance between nearest neighbor VF, when the VF are moving along a line that is perpendicular to their direction of motion (modified from \cite{sridhar2021geometry}). The VF velocity vectors were taken to be identical, so that the VF move in synchrony, and describes a series of burst-and-coast dynamics.}
\label{nemo}
\end{figure}

Within our model the direction in which the RF is moving is dictated by the neuronal firing, which is represented by the ``on" spins in the Ising-like model \cite{sridhar2021geometry}. Each target is described by a group of spins; each one can be either ``on'' (1) or ``off'' (0), such that the net unit vector of the RF's desired direction of motion is given by
\begin{equation}
    \hat{V}= \frac{\sum_{i=1}^m n_i\hat{n}_i}{ \lvert \sum_{i=1}^m n_i\hat{n}_i \rvert}
    \label{V}
\end{equation}
where, $n_i$ refers to the number of ``on" spins in subgroup $i$, $\hat{n}_i$ is the unit vector pointing from the RF towards target $i$ (Fig.~\ref{nemo}), and $m$ is the number of VF (and spin groups). 

Each neural group vector $\hat{n}_i$ is directed, on average, at each of the different VF, but we consider that there is angular noise affecting the angular direction of this vector according to the following Langevin equation
\begin{equation}
    \frac{d\theta_i}{dt}=-\gamma\left(\theta_i(t)-\arcsin\left(\frac{y_i(t)}{\sqrt{(x_i(t)^2+y_i(t)^2}}\right)\right) +\Gamma
    \label{difftheta}
\end{equation}
where $x_i, y_i$ are the coordinates of target $i$ with respect to the location of the RF, $\Gamma$ is the amplitude of the Gaussian noise (with a standard deviation of $\sigma=\frac{\pi}{3}$ and limits of $-b<\Gamma<b$, where $b=\pi$), and $\gamma$ is the rate at which the angle of the internal vector re-adjusts to the new perspective angle of target $i$.

The dynamics of the spins in each group are calculated using the Ising model, based on the following Hamiltonian \citep{n5,sridhar2021geometry}
\begin{equation}
H=-\frac{1}{N}\sum{J_{ij}\sigma_i\sigma_j}
\label{H}
\end{equation}
where $\sigma_i$ are the spin variables, $N$ is the total number of spins in all subgroups, and $J_{ij}$ are the interactions between the spins, given as the dot products of the preferred directions of spins $i$ and $j$, such that: $J_{ij} = J_0 cos(\theta_{ij}^*)$, where $J_0$ is the interaction strength energy, and $\theta_{ij}^*$ is the modified form of the relative angle according to:
\begin{equation}
    \theta_{ij} \xrightarrow[]{}\theta_{ij}^*=\pi\left(\frac{|\theta_{ij}|}{\pi}\right)^\nu
    \label{modifing}
\end{equation}
where $\theta_{ij}$ is the Euclidean  relative angle (Fig.~\ref{nemo}) and the tuning parameter $\nu$ describes the non-Euclidean transformation of the relative angle, thereby determining the interactions between the spins (Fig. S1). Comparisons with experiments indicate that a suitable tuning parameter is $\nu=0.5$ \cite{sridhar2021geometry}, across taxa, which is the value we used in our simulations. 
The dynamics of the spins follow from the Hamiltonian (Eq.~\ref{H}). They can be calculated for discrete, stochastic spins, or using a continuum description in terms of the following Langevin equation for the dynamics of the neural firing
\begin{equation}
\frac{dn_i}{dt} = \left(\frac{N}{m}-n_i\right)k_{i, on} - n_i k_{i, off} + \omega\sqrt{B(n_i)} 
\label{Eqdn_noisy}
\end{equation}
where $B(n_i)$ is the noise amplitude, $m$ is the number of neuronal groups, and $\omega $ is a normally distributed white Gaussian noise (with a standard deviation of $1$). This dynamic is applied to each spin group that represents a target. The noise amplitude $B(n_i)$ is given by \cite{n5} (see SI section S2)
\begin{equation}
B(n_i)=\frac{k_{i, on} (\frac{N}{m}-n_i)+k_{i, off} n_i}{\left(\frac{N}{k}\right)^2}
\label{Bnni}
\end{equation}
where the spin-flip rates (per spin) are given by
\begin{equation}
    \begin{aligned}
    k_{i,off}= k_0\frac{1}{1+\exp\left(\frac{n_i+\sum_{j\neq i}^m n_j\cos(\theta_{ij}^*)}{T}\right)} \\
    k_{i,on}= k_0\frac{1}{1+\exp\left(\frac{-(n_i+\sum_{j\neq i}^m n_j\cos(\theta_{ij}^*))}{T}\right)}
    \label{ki}
    \end{aligned}
\end{equation}
where the sum goes over all spin groups except $i$, $k_0=1 [s^{-1}]$ is a spin-flip rate constant, and $T$ stands for the dimensionless ratio between the noise amplitude and the spin-spin interaction strength, which determines the dynamics of the spins flips. These spin flips represent the dynamics of neuronal switching between firing and non-firing states. Integrating Eq.~\ref{Eqdn_noisy} over time gives us the dynamics of the spins that are ``on'', which determines the direction of the RF motion (Eq.~\ref{V}), as a function of time.

The contribution of each spin group to the desired direction in Eq.~\ref{V} is modified if there are several targets that are very close to each other. Previous work suggests that targets that are within a small relative angle with respect to the animal, are considered as a single target \cite{horn1975mechanism}. To account for this effect, we incorporate an `overlap' function in our model \cite{sridhar2021geometry}. This function reduces the weight of spins encoding a target in Eq.~\ref{V}, if there are other targets in directional proximity (see SI section S3). This effect becomes crucially important for the case of 3VF, as described below (Fig.~\ref{3VF}).

Note that the distance to the target does not influence an individual's direction in the current model: spins that represent proximal targets do not have a larger weight in Eq.~\ref{V} compared to spins that represent more distant targets. We can add such modifications in the future, as needed.

\subsection{Speed of motion: Burst-and-coast dynamics}

In this work, we will consider two possible models for the amplitude of the velocity ($V$), and in both models, this amplitude depends on the distance ($r$) between the RF and the chased VF. Recent experiments demonstrated that the effective interactions between leader-follower have a spring-like property \cite{katz2011inferring,laan2017signatures,zienkiewicz2018data}, whereby the follower increases its speed as its distance to the leader increases, up to a critical distance where it declines. This behavior was also measured for the RF following the VF (see Fig. S4). Note that the precise implementation of this tendency, which keeps the RF from losing the VF that it is chasing, is not the main focus of our model. The novel feature that we introduce relates to directional decision making (as described in the previous section). Therefore, there can be different implementations of the RF speed, as we discuss below. 

In the SI section S5, we give the results for a model where the speed of the RF ($V_{RF}$) is described as a continuous function of the distance between the RF and VF. However, we wish to focus here on a more realistic description of the RF speed, as arising from burst-and-coast dynamics \cite{calovi2018disentangling,li2021burst}, in which the RF's velocity will be given by bursts of force from the tail-beat events, occurring in a stochastic manner. The differential form of the velocity can be written as
\begin{equation}
    \frac{dV_{RF}}{dt}= -\eta V_{RF} +Ff
    \label{V_tail}
\end{equation}
where $\eta$ is the friction coefficient with the water, $F$ is the stochastic force parameter, which is either ``on'' ($F=1$) or ``off'' ($F=0$), and $f$ is the force amplitude of each burst (in units of force/mass). We assume that the rate at which fish exhibit bursts of tail beats (which account for the rapid acceleration in the burst-phase of their motion), $k_{s,on}$, depends solely on the distance of the RF from its targets ($r=|\vec{r}|$, Fig.~\ref{nemo}), and therefore we use the following expression
\begin{equation}
    k_{s,on} (r)=kr\exp{\left(\frac{-r^2}{2r_d^2}\right)}
    \label{Ks_on}
\end{equation}
where $k$ is an effective spring constant ($k=250 [m\cdot s]^{-1}$), and $r_d$ is the length-scale beyond which the fish loses its target (we fit it to be $r_d\sim0.2[m]$). The linear dependence of the burst rate on the distance to the VF, for small values of $r$ (Eq.~\ref{Ks_on}), is motivated by the interactions found between fish \cite{katz2011inferring}, and allows the RF to successfully pursue the VF without losing it. We tested this relation by statistical analysis of the experimental RF's dynamics, which was not conclusive (SI section S7, Fig. S6).

We use this rate in a Gillespie method (see SI section S5) within our Langevin simulations, and it determines when the next burst of tail beats occurs. Following each burst, $F$ is turned ``off'' after a fixed duration, $t_{off}=0.15 [s]$, which for simplicity was chosen to be constant in all the simulations. When the burst of tail beats stops ($F:1\rightarrow 0$), the coasting phase starts, and the velocity of the RF decreases exponentially with time (Eq.~\ref{V_tail}), until the next burst event.

The heading direction of the simulated RF is updated at the onset of each tail burst, since in experiments it was found that the orientation of the RF stays approximately constant for the coasting duration (and is given by Eq.~\ref{V}) \citep{harpaz2017discrete,calovi2018disentangling}.

There are two additional ingredients that we found to be important to implement in our model: In order to better resemble the velocity pattern of the experimental RF, we had to define a velocity threshold ($V_{threshold}$), whereby tail bursting occurs only when the simulated RF's velocity is lower than this threshold (SI section S8). In Fig. S7, we compare the experimental and model RF's velocity dynamic with and without the velocity threshold, of $V_{threshold}=0.04 [m/s]$. The velocity threshold was estimated by analyzing statistical data from the experimental velocity pattern (Fig. S8). 

In addition, we added some noise to the  amplitude of the burst force $f$, as this noise is observed in the experiments (Fig. S9A-D). Hence, we used in our calculations (Eq.~\ref{V_tail}) force amplitudes ($f$) that are drawn from a Gaussian distribution, centered at $f_0$, with a constant variance $\psi=0.2 [m][s^{-2}]$. Note that the experimental RF exerts stronger and faster tail bursts as it is chasing a faster VF (Fig. S9E,F). However, for simplicity the simulated RF in our model is only increasing its burst rate in order to chase a faster VF. As we noted before, our main aim was to explore here the implementation of our spin model for directional decision making, so for the modelling of the RF speed, we merely wanted to implement a reasonably realistic model which can be further improved in the future.

The only parameter that we needed to change as a function of the number of VF that the RF is chasing, in order to obtain reasonable agreements with the experimental observations, was found to be $f_0$: $f_0=1.1 [m][s^{-2}]$ for 1VF, $f_0=1.2 [m][s^{-2}]$ in 2VF, and $f_0=0.95 [m][s^{-2}]$ in 3VF case. It is possible that the RF exerts stronger average tail burst forces when chasing a smaller group of fish since it feels more vulnerable and exposed. Weaker interactions between fish were indeed observed in experiments with larger groups \cite{katz2011inferring}. By reasonable agreement we mean that the simulated and experimental distributions of different quantities that characterize the motion of the RF have the same shape and qualitative features, as well as close average values (see for example the case of 1VF in Fig.~\ref{oneVF}B-D).

When the RF is following several VF, we need to define how the distance $r$ that enters Eq.~\ref{Ks_on} is calculated. The problem can be framed in terms of which VF the RF pays attention to, at each instant. We define a threshold ($\tau$) of fractions of "on" spins (representing the strength of the corresponding neural firing) to determine which targets contribute to the calculation of the distance to targets \cite{sridhar2021geometry}. On each iteration, the level of neural firing that represents each VF ($n_i$) is monitored to determine if it is above the threshold. If so, the distance to the relevant VF is included in the calculation of $r$ as follows
\begin{equation}
    |r|=\frac{\sum_{i=1}^{m} |\vec{r}_i|}{m^*}, \quad \text{for all}\ \frac{n_i}{N/m}> \tau
\label{weighted_r}
\end{equation}
where $m^*\le m$ is the number of neural groups which fire above the threshold, and $|\vec{r}_i|$ is the distance of the RF to the $i$'th VF (Fig.~\ref{nemo}). We used a threshold value of $\tau=10\%$, however the results were found not to be very sensitive to the threshold value (Fig. S10). 

The model parameters are summarized in table ~\ref{table}. Their values are in close agreement with those used in other burst-coast models of this fish motion \cite{sbragaglia2022evolutionary}.
\begin{table}[]
\centering
\includegraphics[]{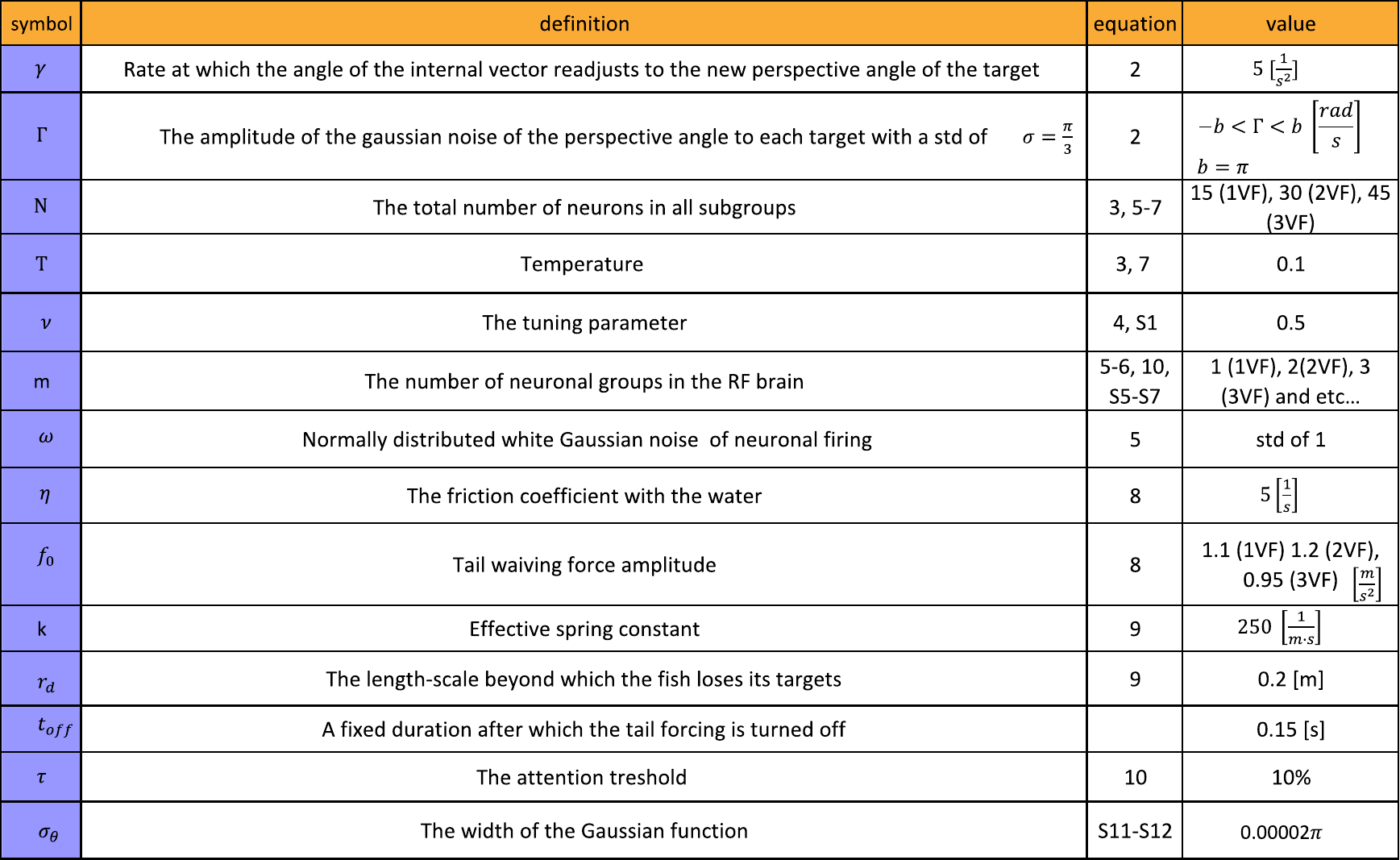}
\caption{Table of parameters used in our model.}
\label{table}
\end{table}

\section{Results}
We compare our model with the experiments in which a RF can school with either one, two, or three VF swimming at different average speeds and spatial configurations. We start with calibrating the model for the case of one VF, and then keep all the model parameters fixed when modeling the behavior of RF following two or three VF, except for slightly modulating the mean burst force $f_0$.

\subsection*{RF following one VF}

\begin{figure}[H]
\centering
\includegraphics[]{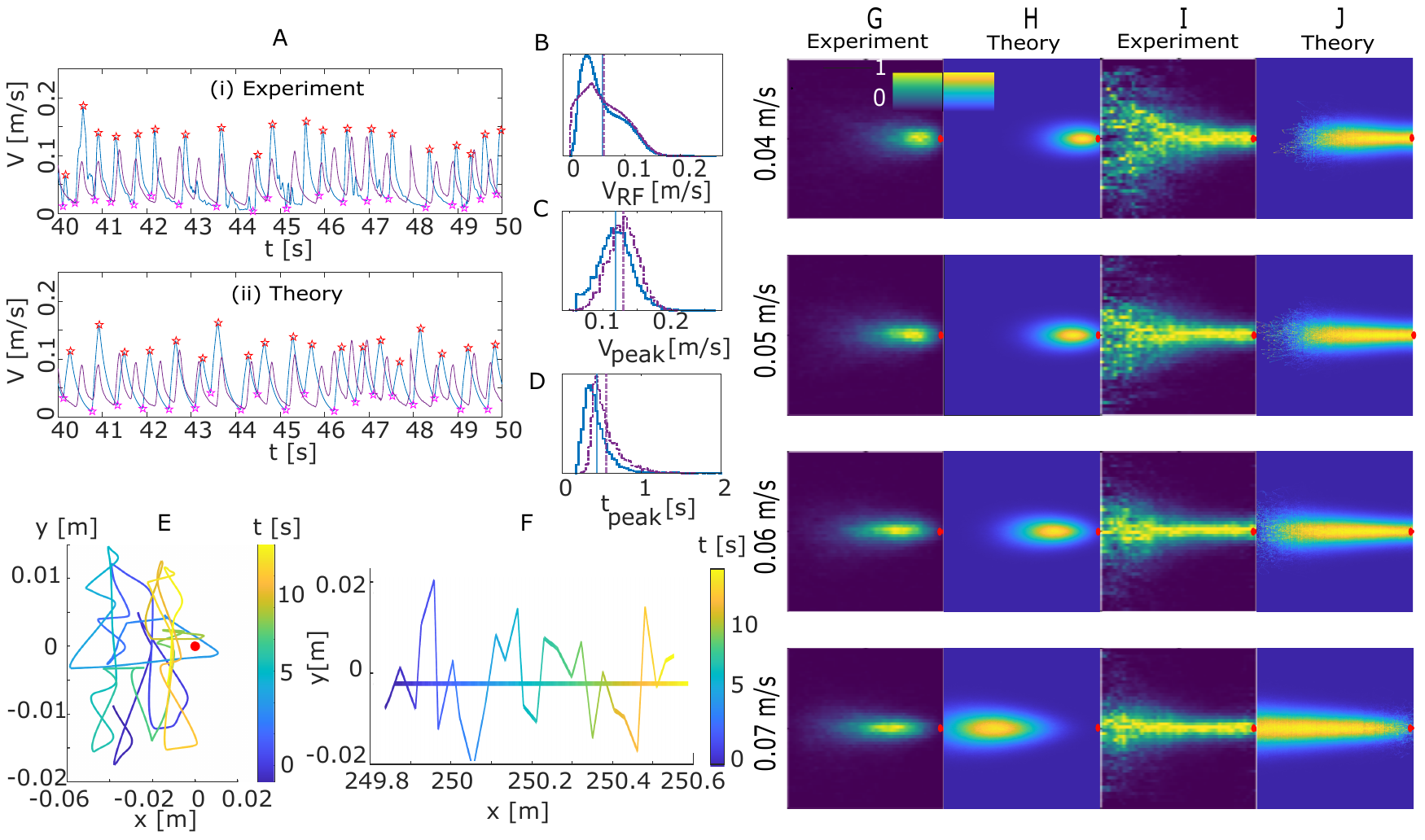}
\caption{RF following one VF: burst-and-coast dynamics and spatial distributions. (A) (i) The RF's speed dynamics in the experiment (blue) compared with the VF velocity (purple, with an average velocity of $V_{VF}=0.05[m/s]$), and (ii) shows the simulated RF velocity (blue) compared with the VF velocity (purple). The maxima and minima of the speed are denoted by the red and pink stars respectively. (B) The speed distribution of the RF, (C) The distribution of the RF's velocity peak values, and (D) The distribution of time intervals between consecutive bursts ($t_{peak}$). In (B-D) the purple line denotes the simulation and blue line the experimental data. (E) An example of the RF trajectory behind 1VF, relative to the VF movement (located at the red dot). (F) Same as (E) but in the lab frame, where the horizontal line gives the trajectory of the VF. In both (E) and (F) time is represented by the color bar. (G-J) Accumulated distribution of the RF behind the VF, for different mean $V_{VF}$ (each row), in the VF frame of reference (the VF is at the origin, red dot). For each velocity, we normalized the heat maps over the whole 2D space (G and H), or over individual $x$-sections (I and J). (G) and (I) show the experimental heat maps, (H) and (J) the simulated results. For each VF velocity, we ran $100$ simulations (in which the RF initial position was random, within a certain distance of the VFs) each for $5,000 [s]$ (500,000 iteration steps).  We used the following parameters: $T=0.1$, $\eta=5 [s^{-1}]$, $\gamma=5 [s^{-1}]$, $\sigma=\pi/3 [rad/s]$ (standard deviation of the angular noise), $b=\pm\pi [rad/s]$ (angular noise limits), $k=250 [(m\cdot s)^{-1}]$, $r_d=0.2 [m]$, $t_{off}=0.15 [s]$, $f_0=1.1[m/s^2]$.}
\label{oneVF}
\end{figure}

We begin with the simplest system of the RF following one VF. A typical time-series of the RF velocity, from the experiment and the simulation, is shown in Fig.~\ref{oneVF}A. The distributions of the RF speed ($V_{RF}$, Fig.~\ref{oneVF}B), maximal speed after each burst ($V_{peak}$, Fig.~\ref{oneVF}C), and the time interval between bursts ($t_{peak}$, Fig.~\ref{oneVF}D), are all in reasonably good agreement between the experiment and simulations. A typical trajectory of the RF chasing the VF is shown in Fig.~\ref{oneVF}E,F, in the VF and lab reference frames, respectively (see Supplementary Movie M1).

The accumulated spatial distributions of the RF behind the VF from our simulations are compared to the experimental observations in Fig.~\ref{oneVF}G-J. The position distributions are shown normalized either over the whole 2D space (Fig.~\ref{oneVF}G,H), or along individual $x$-axis sections (Fig.~\ref{oneVF}I,J). We conclude that our model reasonably fits the experiments, for the different mean speeds of the VF. Further comparisons are shown in Fig. S11.

\subsection*{RF following two VF}

\begin{figure}[H]
\centering
\includegraphics[]{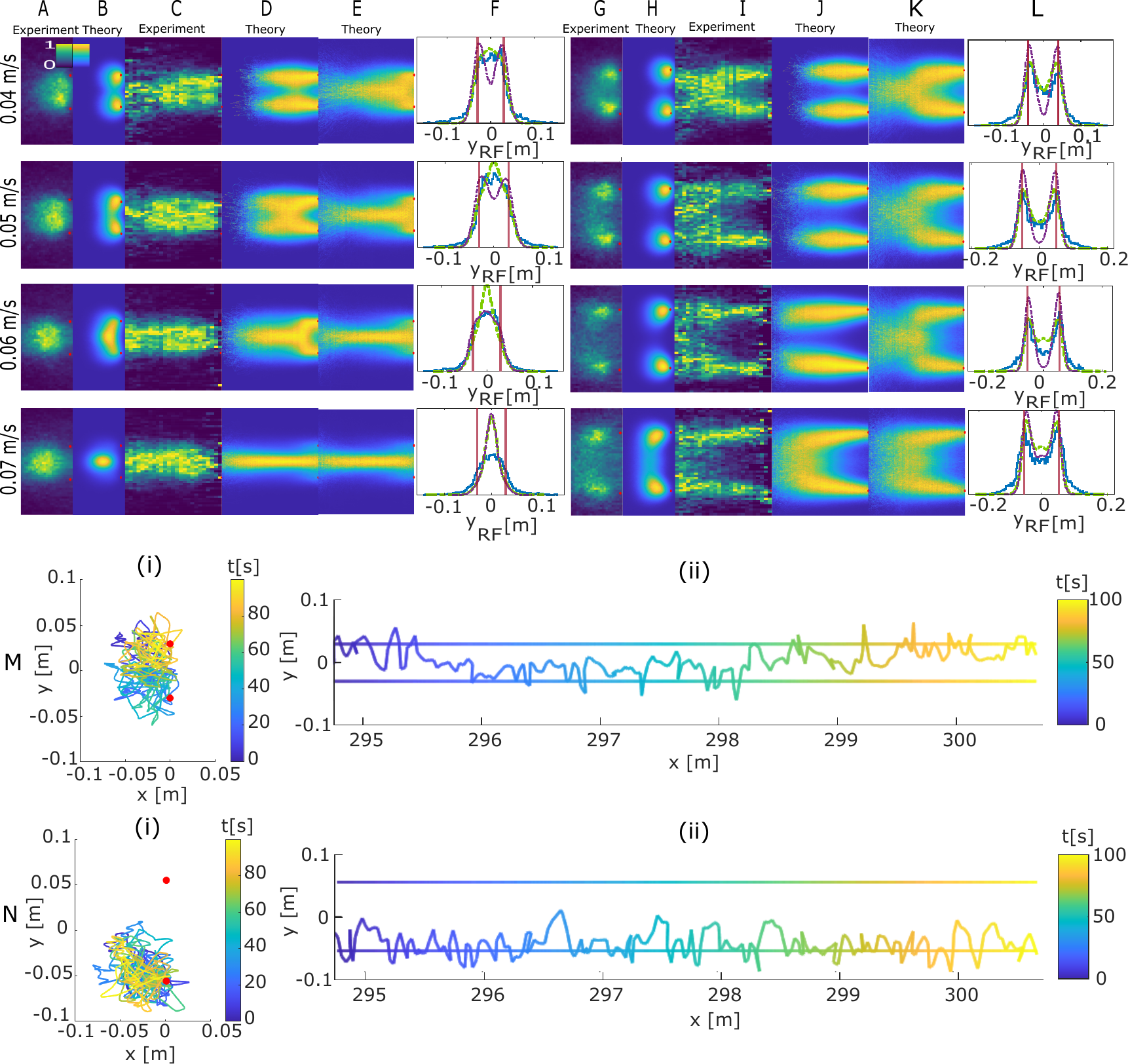}
\caption{Two VF - spatial distributions of the RF, as heat maps normalized over the whole 2D space (A-B, G-H) or over individual $x$-sections (C-E, I-K). (A,G) and (C, I) shows the experimental RF's distribution, while (B, H) and (D, E, J, K) show the corresponding simulated distributions (D,J for long simulations, while E,K for short simulations). (F, L) give the distributions of the RF's $y$-positions and compare the experimental data (blue) with the model (purple for the long simulation and green for the short simulations). The $y$-positions of the VF are denoted by the vertical Maroon lines. (A-F) show results for the compromise regime, in which $LRD=0.06 m$ for all the VF velocities. (G-L) show the results from the bifurcated regime, in which the LRD between targets is chosen to be $0.08[m]$ for $V_{VF}=0.04[m/s]$, $0.1[m]$ for $V_{VF}=0.05[m/s]$, and $0.11[m]$ for $V_{VF}=0.06 [m/s],0.07[m/s])$. We used the identical parameters as in the 1VF system (Fig.~\ref{oneVF}), except for: $f_0=1.2[m/s^2]$. We also added the following parameters (for the neuronal tuning and the overlap function): $\nu=0.5$ and $\sigma_\theta=0.00002\pi$. The long simulations statistics contain $100$ simulations, each ran for $5000[s]$, and the short simulations statistics contains $20,000$ simulations, each ran for $7.5 [s]$. (M) and (N) show typical trajectory of the RF (in a long simulation) in the case of the compromise regime (M), and the bifurcated regime (N), when the VF velocity is $0.06 [m/s]$. M(i) and N(i) show the RF movement relative to the VF, and M(ii) and N(ii) show the movement in the lab frame. For the short simulation in the case of $0.07 [m/s]$ the initial condition of $x_{RF}$ is randomized in the range of $0 -(-0.06) [m]$ behind the VF, while in all other plots it is randomized in the range of $0-(-0.1) [m]$.}
\label{twoVF}
\end{figure}

In Fig.~\ref{twoVF} we plot the accumulated spatial distribution of the RF chasing two VF that are moving at different average speeds, and at different lateral separations (left-right distance, LRD, Fig.\ref{nemo}). In Fig.~\ref{twoVF}A-F we plot the results for a small value of LRD ($=0.06[m]$), where the experimental data show the RF to be mainly positioned behind, and in-between the 2VF, and no clear bifurcation is observable in the RF distributions (Fig.~\ref{twoVF}A,C). At larger LRD values (Fig.~\ref{twoVF}G-L), we clearly see that when the RF is close to the two VF, it follows either one of them, while when it is further away it is found more in the compromise position between them (Fig.~\ref{twoVF}G,I). Further comparisons are shown in Fig. S12.

In our model, the motion of the RF between the two VF is termed the ``compromise'' regime, where the spin dynamics do not break the symmetry. When the symmetry is broken, one group dominates, and the other is inhibited. Such a bifurcation, termed ``decision'', arises in our model when the relative angle between the targets, with respect to the RF, is larger than a critical value (of $\sim 90^{o}$ for our choice of model parameters) \cite{sridhar2021geometry}.

Comparing to the simulations, we find that for the small LRD value and low VF speeds ($V_{VF} =0.04,0.05,0.06 [m/s]$), the RF distribution is already bifurcated, i.e. the RF follows behind each of the VF, and not in the compromise position between them (Fig.~\ref{twoVF}B,D, and Supplementary Movie M2). This behavior is not observed in the experiments, and arises from the fact that the experimental trajectories of the VF and the RF are bounded by the container walls, thereby limiting the experimental data to short trajectories (of $\sim6-8$s duration) that are interrupted by U-turns, while the simulated trajectories are long and uninterrupted (around $5,000 [s]$). We demonstrate this in Fig.~\ref{twoVF}E,K, by comparing the RF's spatial distributions obtained from many short simulated trajectories (of $7.5 [s]$ duration), that match better the experimental distributions. During short trajectories the spatial distribution of the simulated RF does not reach its steady-state distribution, as obtained from simulating long trajectories. 

At the larger LRD values, both the experimental and the simulated distributions exhibit clear bifurcation of the RF dynamics, which follows either one of the two VF (Fig.~\ref{twoVF}G-L, and Supplementary Movie M3). The projections of the RF's positions on the $y$-axis (Fig.~\ref{twoVF}F,L), show that there exists good agreement between the model and the experiments. Similarly, the distributions of the speed, velocity peaks, and intervals between consecutive velocity peaks are in reasonable agreement between the model and the experiments (Fig. S12). 

Note that the overlap function did not significantly affect the RF distribution for the simulated two VF system, except at the fastest velocity of the VF ($0.07 [m/s]$). This is expected, as the overlap affects the dynamics when the RF is far behind the two VF, so that their relative angle is smaller, which occurs more for the fastest moving VF. At this velocity, without the overlap function, the bifurcation event occurred too close to the targets.

\subsection*{RF following three VF}

\begin{figure} [H]
\centering
\includegraphics[]{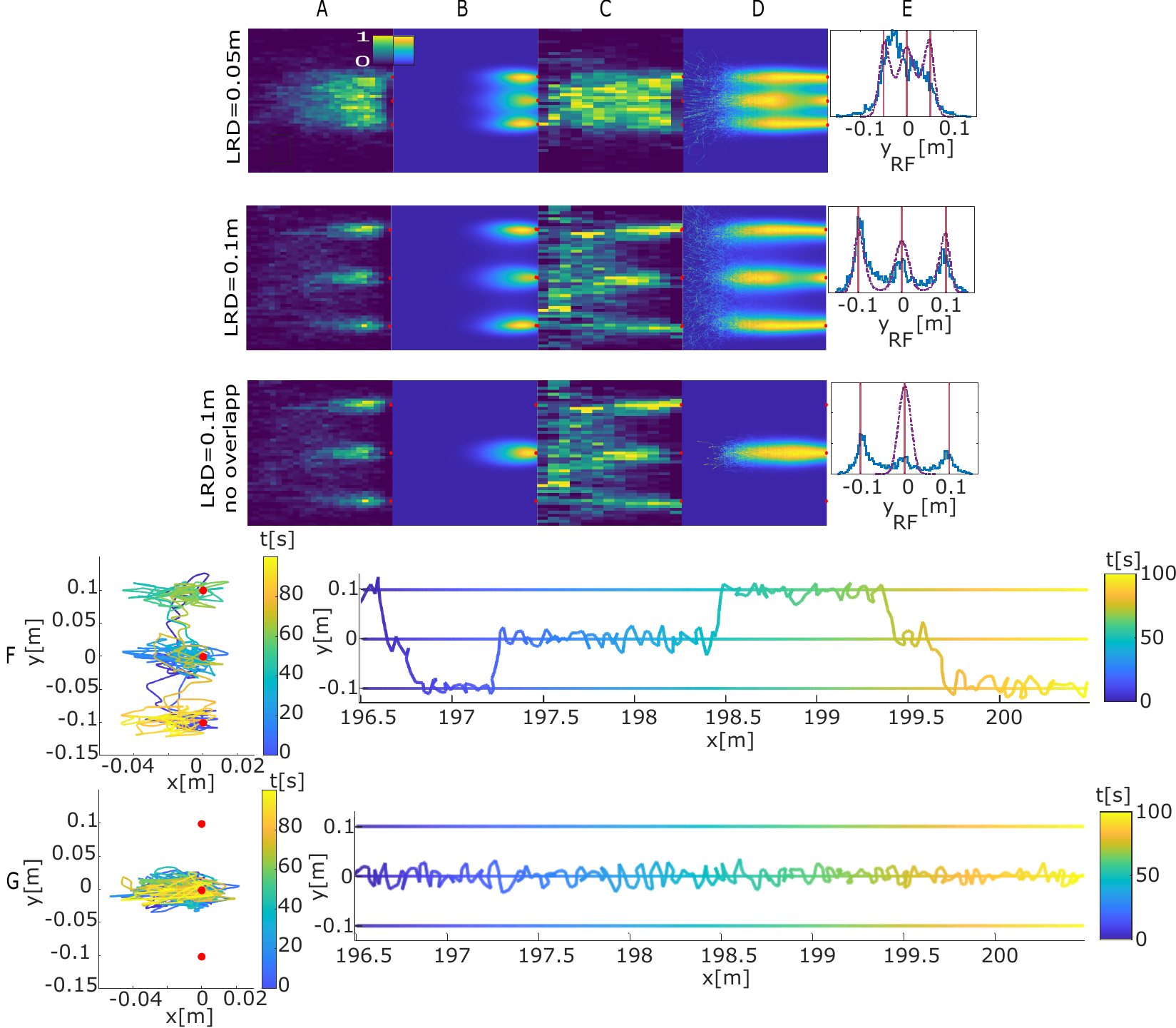}
\caption{RF following 3VF with an average velocity of $V_{VF}=0.04[m/s]$. The  value for each line in (A-E) is given on the left, with the third-panel presenting simulations without the overlap function. The heat maps present the accumulated spatial RF distribution, normalized over the whole 2D space (A,B) or over the $x$-axis sections (C,D). (A,C) Experimental data, while (B,D) the simulation results. (E) Distributions of the RF's projected $y$-positions (experiments in blue and simulations in purple). The $y$-positions of the VF are denoted by the vertical Maroon lines. (F,G) Examples of simulated RF trajectories behind the 3VF when $LRD=0.1 [m]$ with (F) and without (G) the overlap function. The left panels show the trajectory relative to the 3VF (red dots), and the right panels show the trajectory in the lab frame (trajectories of the 3VF are given by the horizontal lines). Time is given by the color bar. For 3VF cases, we used the same parameters as in 2VF (Fig.~\ref{twoVF}), except for the burst force amplitude which was changed to $f_0=0.95 [m/s^2]$. we ran $100$ simulations (in which the RF initial position was random) each for $5000 [s]$ (500,000 iteration steps).}
  \label{3VF}
\end{figure}

The experiments and the simulation results for three VF are displayed in Fig.~\ref{3VF}. We find an overall good agreement between the accumulated spatial distributions from the experiments and the corresponding simulations (Fig.~\ref{3VF}A-E top two lines). The bifurcations are clearly observed in the model and in the experiments for the larger LRD value ($LRD=0.10[m]$), while for $LRD=0.05[m]$ we see the bifurcations clearly only in the simulations but not in the experiments. This is likely due to the larger experimental noise, and to the shorter trajectories in the experiments compared to the long trajectories used in these simulations (as in the 2VF case, see Fig.~\ref{twoVF}E,K). Further comparisons are shown in Fig. S13, and in Supplementary Movies M4-M5.

The overlap function makes a crucial difference in the dynamics of the RF in the system with three VF, as we demonstrate by plotting the simulated distribution when this effect is not taken into account (Fig.~\ref{3VF}A-E bottom line). Without the overlap function, the RF predominantly follows the middle target due to geometrical reasons. Following the middle target appears to be a steady-state solution in the RF decision-making process, in the absence of the overlap normalization, as demonstrated by the sample trajectory (Fig.~\ref{3VF}G). In the presence of the overlap function, we observed that the RF chases behind each of the three VF and switches between them (Fig.~\ref{3VF}F), and we could also distinguish the bifurcation events on the density maps (Fig.~\ref{3VF}A-D, $LRD=0.1[m]$), in agreement with the experimental observations. For small LRD values, the effects of the overlap function are not visible (Fig. S13).

\subsection*{2VF in shifted configuration} \label{Assy}
\begin{figure}[H]
\centering
\includegraphics[]{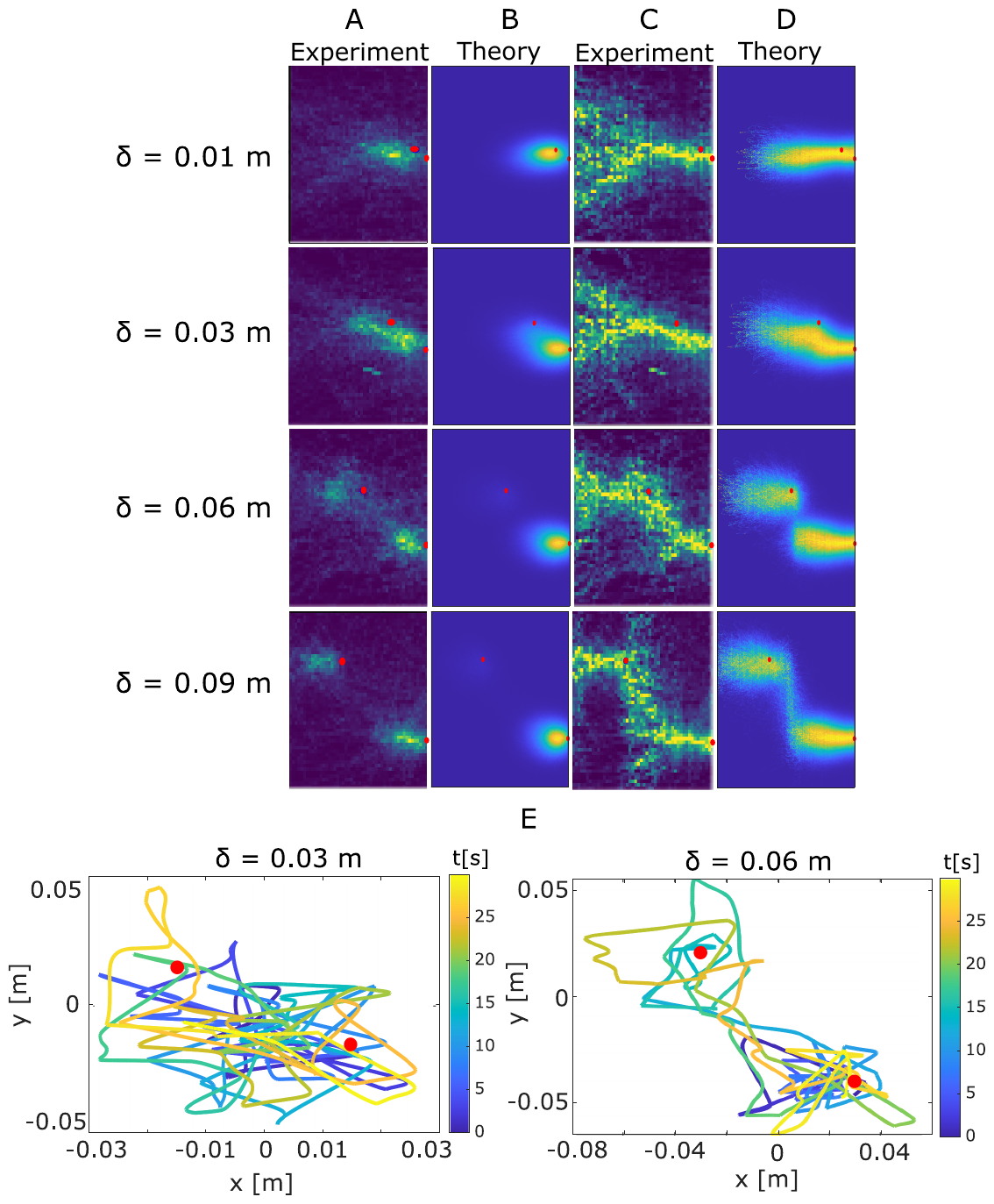}
\caption{2VF in a shifted geometry: one of the VF is shifted in the front-back and the left-right directions with respect to the other VF, by a distance of $\delta$. Both of the VF swim with an average velocity of $V_{VF}=0.04[m/s]$. The accumulated spatial distribution of the RF behind the 2VF, (A,B) normalized over the whole 2D space, or (C,D) normalized over $x$-axis sections. (A) and (C) show the results from experimental data, while (B) and (D) show results from the simulations. The VF are denoted by the red circles. The model parameters are the same as those used in Fig.~\ref{twoVF}. (E) Examples of RF trajectories relative to the 2VF (red dots), for the cases of $\delta=0.03 [m]$, and $\delta=0.06 [m]$. In both plots, the color of the line represents the time, as indicated by the colorbar, indicating that the RF can go back to follow the trailing VF.  }
  \label{Symy}
\end{figure}

Next, we investigated a configuration of 2VF which are not aligned along the $y$-axis, i.e. along a line that is perpendicular to their direction of motion. One VF was shifted in both the front-back and the left-right directions with respect to the other VF, by a distance of $\delta$. The accumulated spatial distributions from both the experiments and simulations are compared in Fig.~\ref{Symy}A-D, for different values of the displacement. Overall the model results match the experimental distributions. For example, even in the `bifurcation' regime ($\delta=0.06[m]$ in Fig.~\ref{Symy}A-D), both the experiment and model show that the RF moves between both VF. Typical simulated trajectories demonstrate the motion of the RF between the two VF in Fig.~\ref{Symy}E. Further comparisons are shown in Fig. S14, and in Supplementary Movie M6.

This agreement demonstrates that our model is not limited to describing a fish responding to a single line of leaders, but can also account for the fish motion within a moving group with complex geometry. Note that the one-dimensional version of the model, used in \cite{sridhar2021geometry}, cannot explain the trajectories in this configuration.

\section{Discussion}

Our research provides a novel approach to modeling the motion of fish when responding socially to other fish, by including an explicit model for the directional decision-making processes of the individual fish. This is implemented using a spin-based model \cite{sridhar2021geometry}, which gives rise to spontaneous symmetry breaking, describing the cognitive process of choosing which of the targets to follow. Previous models of animals following others moving in a group usually rely on alignment interactions or vectorial averaging, which do not contain a mechanism for choice. While the parameters used in our model simulations are specific to the zebrafish (\textit{Danio rerio}) that were examined in the experiments, the proposed directional decision-making model may apply more generally to other animals that move in groups. 
 
The Ising spin model that we used for the directional decision-making, is shown to provide an appropriate framework for representing the fish's directional decisions while it is chasing moving targets. While previously applied under the constraints of projecting it to a 1D motion \cite{sridhar2021geometry}, in this work we have extended it to describe the free movement of the fish in 2D when following other moving conspecifics. The model simulations recovered the main dynamical features of the experimentally observed RF motion. Most importantly, the model captures the observed bifurcation transition exhibited by real fish, as a function of the separation between the VF: when the VF are close to each other, the RF follows them along an averaged (compromise) position, while when the VF are far apart, the RF pursues each of them individually. Moreover, we used identical parameters for VF of different speeds and geometries, indicating that the model is rather general and robust. 
 
Our model may form a first step towards implementing individual directional decision-making in models of animal collective motion. As shown in Fig. S15, we can explore the dynamics of many RF interacting using our model, following a single leader, or in its absence. These results indicate that the model can describe the formation of cohesive shoal (``swarm'') behavior, without any spontaneous alignment of the fish along a particular direction \cite{delcourt2012shoals} (see Supplementary Movie M10). This suggests that for the maintenance of collective persistent motion along a particular direction (``schooling phase'' \cite{2013collective,wang2022impact}), the spontaneous emergence of leaders is needed \cite{conradt2009leading,zienkiewicz2015leadership,mwaffo2018detecting}, such as leadership by indifference (whereby less socially-responsive individuals spontaneously become leaders \cite{conradt2009leading}), which is not included in the current model.  Note that our model ignores the finite size of the RF, which means that we do not describe the short-range maneuvers that are necessary for collision avoidance. Similarly, in a real fish school some individuals may have more influence than others \cite{butail2016model,heras2019deep}. In addition, our model explicitly involves many-body interactions, that are not simply the superposition of pairwise interactions. Significant interactions beyond pairwise have indeed been measured inside fish schools \cite{katz2011inferring}.

The model can be elaborated and developed to include more detailed aspects of the interactions between the fish \cite{butail2016model,zienkiewicz2018data,heras2019deep,lei2020,li2020vortex,wang2022impact}, while the explicit directional decision-making model may be incorporated into existing models of collective animal motion \cite{rahmani2020flocking,bastien2020model}. It remains for future exploration how this individual decision-making process can give rise to the emergence of polarized fish schools and (possibly transient) leaders.

\section{Acknowledgments}
We thank Pawel Romanczuk for useful comments. N.S.G. is the incumbent of the Lee and William Abramowitz Professorial Chair of Biophysics. This research is made possible in part by the historic generosity of the Harold Perlman Family. D.G. and I.D.C. acknowledge support from the Office of Naval Research Grant N0001419-1-2556, Germany’s Excellence Strategy-EXC 2117–422037984 (to I.D.C.) and the Max Planck Society, as well as the European Union’s Horizon 2020 research and innovation programme under the Marie Sk{\l}odowska-Curie grant agreement (to I.D.C.; $\#$860949).

\bibliographystyle{apsrev4-1}
\bibliography{paper}

\end{document}


\newcommand{\be}{\begin{equation}}
\newcommand{\ee}{\end{equation}}
\newcommand{\bea}{\begin{eqnarray}}
\newcommand{\eea}{\end{eqnarray}}
\newcommand{\nn}{\nonumber}

\title{A simple cognitive model explains movement decisions during schooling in zebrafish: Supplementary Information}
\author{Lital Oscar$^1$, Liang Li$^{2,3,4}$, Dan Gorbonos$^{2,3,4}$, Iain D. Couzin$^{2,3,4}$ and Nir S. Gov$^1$\footnote{nir.gov@weizmann.ac.il}}
\affiliation{$^1$Department of Chemical and Biological Physics, Weizmann Institute of Science, Rehovot 7610001, Israel \\ $^2$Department of Collective Behaviour, Max Planck Institute of Animal Behavior, 78464 Konstanz, Germany\\ $^3$Centre for the Advanced Study of Collective
Behaviour, University of Konstanz, 78464 Konstanz, Germany\\ $^4$Department of Biology, University of Konstanz, 78464 Konstanz, Germany}
\maketitle
\renewcommand{\thesection}{S\arabic{section}}
\renewcommand{\thefigure}{S\arabic{figure}}
\setcounter{figure}{0}
\renewcommand{\thetable}{S\arabic{table}}
\renewcommand{\theequation}{S\arabic{equation}}

\setcounter{table}{0}
\setcounter{equation}{0}

\section{Tuning parameter of the relative angles}\label{Overlapfunction}
\begin{figure}[H]
    \centering
    \includegraphics[scale=1]{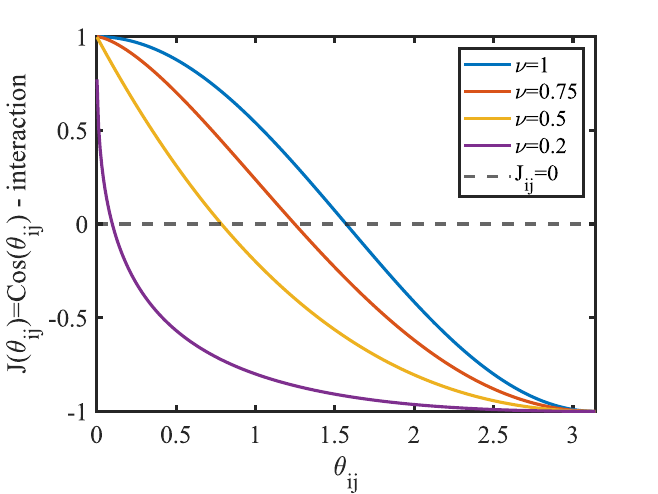}
    \caption{\footnotesize{The tuning parameter affect the conflict angle between targets ($\theta_{ij}$) in which we get inhibition ($J_{ij}<0$) or excitation ($J_{ij}>0$). As the tuning parameter is smaller, the conflict angle needed to inhibition is smaller. We used $\nu=0.5$ (in orange) in our fish model.}}
    \label{tuning}
\end{figure}

We investigated a modified form of the neuronal interactions based on the angle between its targets. The interaction between the spins of different sub-groups uses a modified angle according to:
\begin{equation}
    \theta_{ij} \xrightarrow[]{}\theta_{ij}^*=\pi\left(\frac{|\theta_{ij}|}{\pi}\right)^\nu
    \label{modifing}
\end{equation}
where $\theta_{ij}$ is the original measured conflict angle (as shown in Fig.1), $\theta_{ij}^*$ is the new modified angle and the tuning parameter $\nu$ describes the shape of the interaction between neural circuits, corresponding to the wiring or filtering activity of the brain's neurons (Fig.\ref{tuning}). Hence the interaction between the spins was modified to:

\begin{equation}
    J_{ij}=cos(\theta_{ij}^*)
    \label{cos_mod}
\end{equation}
From previous experiments, a suitable tuning parameter to represent animal movement was found to be $\nu{\approx}0.5$ \cite{sridhar2021geometry}, therefore we will use this value in our simulations. In Fig.\ref{tuning} we plot the affect of $\nu$ on the conflict angle in which excitation ($J_{ij}>0$) changes to inhibition ($J_{ij}<0$).

\section{Calculating the neuronal noise term in the continuum model}

To calculate the neural noise which we added to the continuum approach, for each subgroup of neurons, we first defined $r_-$ and $r_+$ in each group to be:

\begin{equation}
  \begin{array}{l}
    r_+ = k_{i, on} (\frac{N}{m}-n_i) \\
    \\
    r_- = k_{i, off} n_i
    \label{r_-}
    \end{array}
\end{equation}
where, $m$ is the number of neural groups (and also the number of VF), and $n_i$ is the amount of "on" neurons in group $i$. By substituting the Glauber rates as defined in Eq.7 in Eq.\ref{r_-}, we get:

\begin{equation}
  \begin{array}{l}
        r_+ = \frac{\frac{N}{m}-n_i}{1+\exp{\left(-\frac{n_i + \sum_{j\neq i}^m n_j\cos{\theta_{ij}}}{T}\right)}} \\
        \\
        r_- = \frac{n_i}{1+\exp{\left({\frac{n_i + \sum_{j\neq i}^m n_j\cos{\theta_{ij}}}{T}}\right)}}
        \label{r}
\end{array}
\end{equation}

We will focus on one event change at most in $\delta n_i$:
\begin{equation}
    \delta n_i=	  \begin{cases}
    +\frac{1}{\frac{1}{m}N}      & \quad P_+=r_+ \delta t\\
    -\frac{1}{\frac{1}{m}N}   & \quad P_-=r_- \delta t
  \end{cases}
\end{equation}
here $P_+$ and $P_-$ are the probabilities of the neurons to turn "on" or "off, and the variance in neural firing is:
\begin{equation}
    \delta n_i^2=\frac{P^+ +P^-}{\left(\frac{N}{m}\right)^2}=\frac{r_+\delta t+ r_-\delta t}{(\frac{N}{m})^2}
\end{equation}
 Next, we calculate the neural noise amplitude, $B(n_i)$:
\begin{equation}
    B(n_i)=\lim\limits_{\delta t \to 0} \frac{<\delta n_i^2>}{\delta t}
\end{equation}
\raggedbottom
and by substituting $\delta n_i^2$ inside the previous equation, we get:
\begin{equation}
    B(n_i)=\frac{r_++r_-}{\left(\frac{N}{m}\right)^2}
    \label{Bn}
\end{equation}
We will substitute Eq.\ref{r} in Eq.\ref{Bn} and this will be calculated each time step to find the neural noise amplitude which depend on $N$, $T$, $n_i$, $n_j$, and $\theta_{ij}$:

\begin{equation}
B(n_i)=\frac{k_{i, on} (\frac{N}{m}-n_i)+k_{i, off} n_i}{\left(\frac{N}{m}\right)^2}
\label{Bnni}
\end{equation}

We used Eq.\ref{Bnni} as the amplitude of the neural noise as follows:
\begin{equation}
\frac{dn_i}{dt} = \left(\frac{N}{m}-n_i\right)k_{i, on} - n_i k_{i, off} + \omega\sqrt{B(n_i)} 
\label{Eqdn_noisy}
\end{equation}

$B(n_i)$ is the noise amplitude, and $\omega$ is a normally distributed white Gaussian noise (with a standard deviation of $1$ in our model).

\section{The overlap function}\label{Overlapfunction}
\begin{figure} [H]
    \centering
    \includegraphics[scale=0.4]{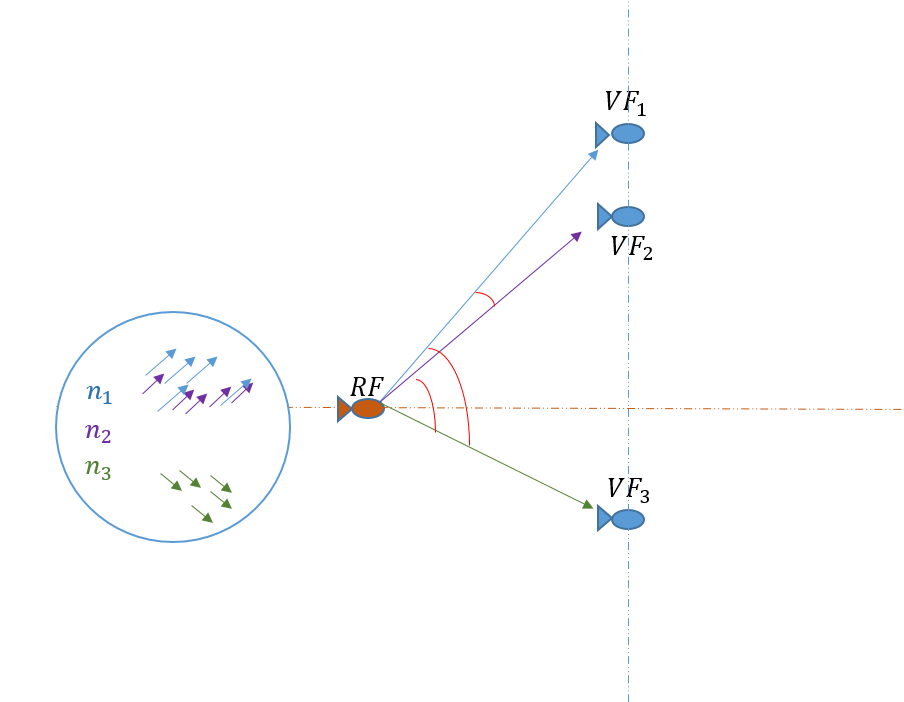}
    \caption{\footnotesize{The Overlap function in cases of overlapping targets -  The RF observes $VF_1$ and $VF_2$ at very similar angles, yet it sees $VF_3$ at a distinct angle. Therefore, the overlap function for the RF has been employed to avoid bias to the direction of the $VF_1$ and $VF_2$. This decreases the total amount of neural firing in $n_1$ and $n_2$ depending on how the related angles overlap (how close is $\theta_1$ to $\theta_2$). The $n_3$ group may also be affected, but to a much lesser degree.}}
    \label{fig:O}
\end{figure}

The main objective for the overlap function is to represent a situation in which several VF are within very close viewing angles with respect to the RF eyes. In such cases, we want the RF to "see" them as "one" target, so the neural groups will not fire fully to both VF, but will overlap with each other. This will decrease the total neurons that can be turned "on" in each group. In this manner, the fish will not tend to follow the overlapping targets more than isolated VF, since it sees them as approximately one fish. 
Figure \ref{fig:O} illustrates such a case in which the overlapping function could affect the neural firing of groups number $1$ and $2$ but does not change the dynamics in group number $3$. The RF sees both $VF_1$ and $VF_2$ at very similar angles, and therefore their relative neural groups will overlap and will be reduced.

\begin{figure} [H]
    \centering
    \includegraphics{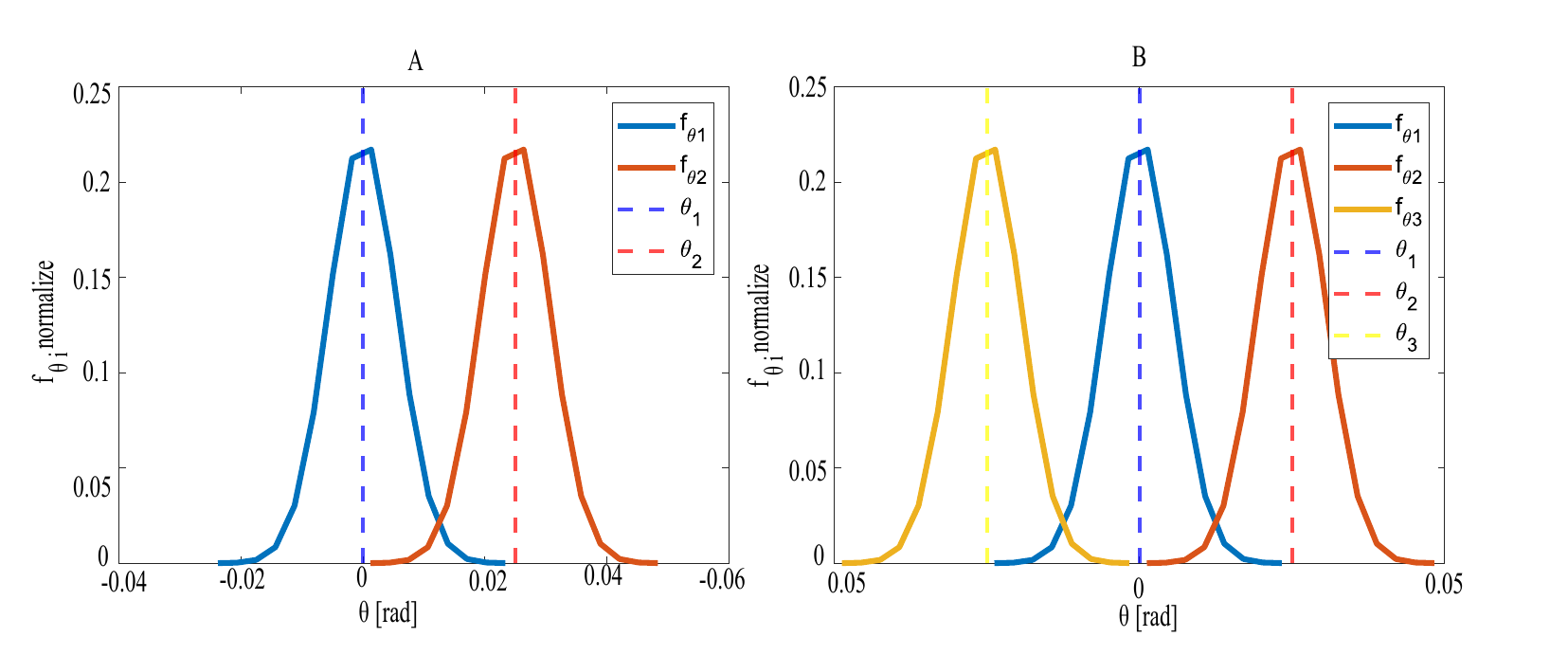}
    \caption{The Gaussian function of Eq.\ref{F} representing the neuronal spread of the neurons that point to each target. The right plot (A) shows two targets configuration (2VF), and the left plot (B) shows three targets configuration (3VF). The 3VF is symmetrical around the angle $\theta_1$ directed at the central target. We chose a value of $\sigma_{\theta}=0.00002\pi$ for both 2VF and 3VF. The values of the normalization factor in these examples are: in 2VF - overlap function of groups $1$ and $2$ get $0.9805$. In 3VF- overlap function of groups $1$ and $2$ is $0.9805$ and the overlap function of group $3$ equals $0.9610$.}
    \label{Oexample}
\end{figure}

Each set of spins ($n_i$) is aimed at angle $\theta_i$ which points towards target $i$, but for simplicity we assume it has a Gaussian form around this angle:

\begin{equation}
f_i(\theta_i, \theta)= A\exp^{-\frac{(\theta-\theta_i)^2}{\sigma_\theta}}
\label{F}
\end{equation}
Where $\sigma_\theta$ is the width of the Gaussian function, and $A$ is a normalization factor such that: $\int_{\theta_i-b}^{\theta_i+b}  f_i(\theta_i, \theta) d\theta =1$ (the limits of the integral $b$, will be given next). To demonstrate the overlap functions, we present Fig.(\ref{Oexample}).

The normalization of the neuronal strength due to the overlap function is calculated in the following manner:

\begin{equation}
n_i=n_{i.0}\int_{\theta_i-b}^{\theta_i+b} \frac{f_i(\theta_i, \theta)}{\sum_{j} f_j(\theta_j, \theta)}f_i(\theta_i, \theta) d\theta
\label{overlapping}
\end{equation}

Here $b=3\sqrt{\sigma_\theta}$, $n_{i,0}$ is the original neural firing before including the overlapping, $f_i(\theta_i, \theta)$ is from Eq.\ref{F} and $f_j(\theta_j, \theta)$ is the same just for other VF, in the direction of $\theta_j$. The sum in the denominator goes over all the subgroups (to each VF), including $j=i$. 
At every angle $\theta_i$ the fraction of the spins of group $i$ that are counted is not $f_i (\theta_i,\theta)d\theta$, but decreased due to the overlap with the other Gaussian functions around the $\theta_j$, as given by the normalizing factor $\frac{f_i(\theta_i, \theta)}{\sum_j f_j(\theta_j, \theta)}$.

In 2VF, when the VF's angles overlap ($\theta_1-\theta_2 \xrightarrow[]{}0$), the function goes to $\frac{1}{2}$, and we do not double-count the spins that point at the same target. When the targets do not overlap, each recieves its full share of $n_{i,0}$ spins.
In 3VF case, there can be symmetry around the middle target ($\theta_1$). When the three targets overlap, the overlap function of each of them goes to $1/3$ so that we do not triple-count the spins that point at the same target. The central target is affected by a stronger overlap since it is affected by both edge targets.
\par
We kept the parameter $\sigma_{\theta}$ equal to $0.00002\pi$ on our simulations. We confirmed that the overlap contributes to our model in Fig.\ref{3VF}. Before appending the overlap function, the RF only chases behind the middle target of the 3VF (Fig.\ref{3VF}). However, when the overlap is included, the RF chases behind all three VF, which better fits the experiment. The reason is that the overlap causes the middle VF to have lower amplitude due to overlapping with the nearby VF from both sides. Subsequently the neural firing of the group directed at the middle target decreases. 

\section{Effective interaction between the RF and VF, experiments and model}\label{Attention}

\begin{figure} [H]
\centering
\includegraphics[]{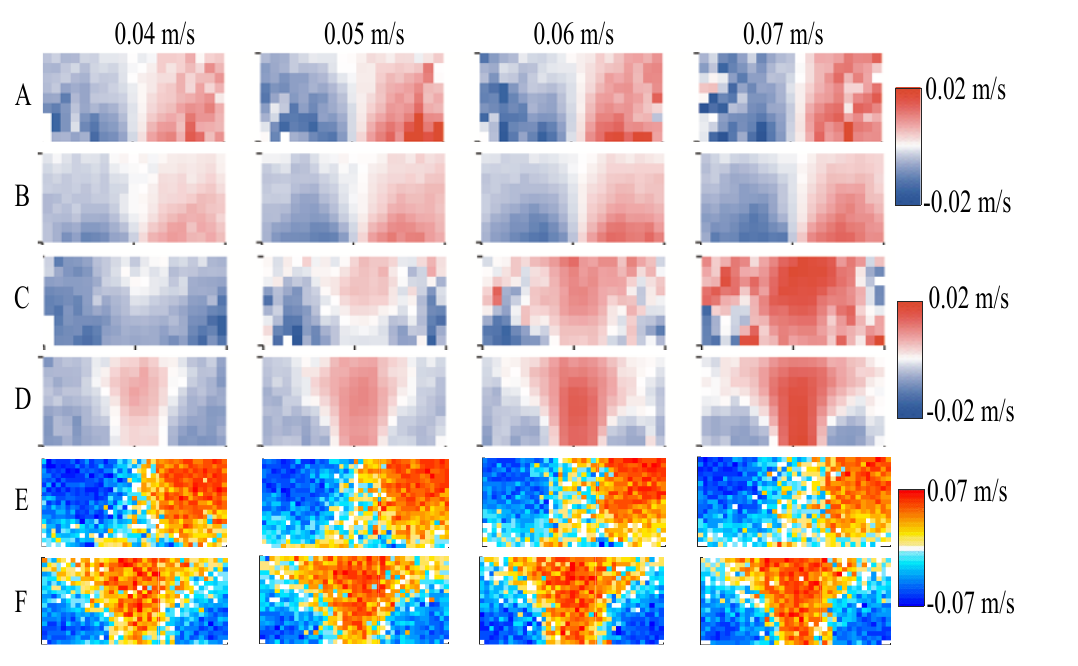}
\caption{The effective interactions between fish, as manifested by their acceleration as function of their position in relation to the fish they are chasing. The chasing RF is at the origin $(0,0)$ and the target is at different positions around it. The $x$-axis is the front back distance between the RF and its target, $y$-axis is the relative left/right displacement. The turning speed and forward speed components are shown by the colormap. (A) Experimental data showing turning components of the RF's speed when chasing another RF, or a VF (B). (C) Experimental data showing forward components of the RF's speed when chasing another RF, or a VF (D). (E-F) The simulated turning (E) and forward speed (F), using the same parameters as in Fig.2.}
\label{fig_speeding}
\end{figure}

The relative acceleration of a fish following other fish, or another VF, was measured in experiments (Fig.\ref{fig_speeding}A-D). A similar analysis of our simulations is shown in Fig.\ref{fig_speeding}E,F.

\section{Spring-like model for the RF speed}
Before applying the tail burst events in our model, we tried a simpler approach, where the velocity amplitude is directly calculated in a continuous manner from the distance of the RF to its targets (depending on the attention of the RF). The RF velocity amplitude is calculated in the following way:

\begin{equation}
    V_0 (|\vec{r}|) = k |\vec{r}|\exp{\left(\frac{-|\vec{r}|^2}{2r_d^2}\right)}
    \label{v0}
\end{equation}
where $\vec{r}=\vec{r}_{rf}-\vec{r}_{vf}$ is the absolute distance of the RF from the VF, $V_0$ is the speed value, $r_d$ is a constant that gives the range beyond which the interactions between the fish decay, and $k$ is the spring constant.  We added a finite time scale that adjusts the speed of the RF in order for it to reach closer to the VF as in experiments. Together with a noise in the velocity of the RF, such that the velocity distribution is wider and fit better. Both were added by applying the following equation:

\begin{equation}
    \frac{dV}{dt}= -\beta(V-V_0) +\xi
    \label{beta}
\end{equation}

where $\beta$ is the frequency to adjust the velocity, and $\xi$ is a white Gaussian noise term.

\begin{figure} [H]
\centering
\includegraphics{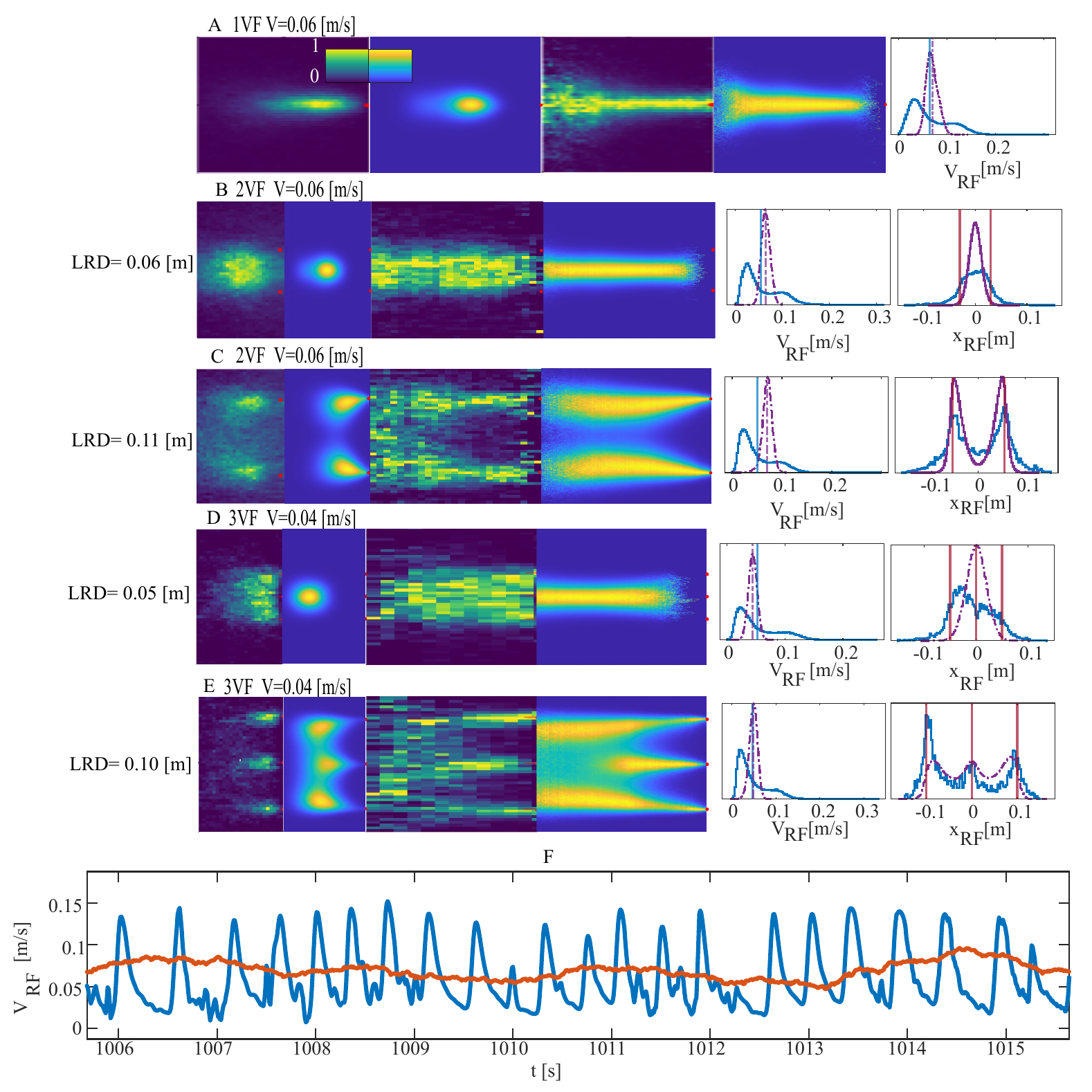}
\caption{Spring Like interactions model - Heat maps of the spatial distribution of the RF chasing (A) 1VF, (B,C) 2VF and (D,E) 3VF. The right panels give the speed, and projected $y$-position distributions. The blue/purple line show the experimental/simulated distribution. The bottom plot show the RF dynamics in experiment (blue), and the red line the RF velocity from the spring-like simulation. For good statistics we conducted for each heat map and histogram $500$ simulations, with random initial position of the RF, each for $5,000 [s]$ which equal to 500,000 iteration steps.}
\label{SI_SL} 
\end{figure}

We apply this spring like method on the the two dimensional system for one, two and three targets (Fig.\ref{SI_SL}) This model resulted in a reasonably good agreement with experimental heat maps (Fig.\ref{SI_SL}A-E). However, the RF's velocity dynamics did not match the velocity pattern of the RF in experiments (Fig.\ref{SI_SL}F). Therefore, we focused in the main text on a model of realistic burst-and-coast behavior of the RF. The spring-like model did not give realistic spatial distributions when the 2VF where in a shifted geometry (Fig.5).

\section{Burst events}\label{Burstevents}
In order to simulate the tail bursts we used a Gillespie method to determine whether a tail burst just occurred or not, depending on the rate of the tail to exert a burst (Eq.9). The Gillespie simulation determines in each iteration whether a tail burst just happen ($F=1$) or stopped ($F=0$). The first step is to determine the time of the next event, a number $r$ is drawn from a uniform distribution and the next time step is calculated by:
\begin{equation}
dt = \frac{1}{k_{s,on}}log\left(\frac{1}{r}\right)
\label{C1}
\end{equation}

where $k_{s,on}$ the rate to burst, as given in Eq.9. We defined the time step between iterations to be  $\Delta t=0.01 [s]$, so that during each iteration we check whether $dt < \Delta t$ and also if $V_{RF}<V_{threshold}$ (see section \ref{Vtreshold} below). If both of these conditions are satisfied,  a tail burst occurs and $F=1$, which is then inserted in Eq.8, and affects the RF's velocity.

\section{The burst rate vs the distance to target}\label{rate_linear}
\begin{figure} [H]
    \centering
    \includegraphics{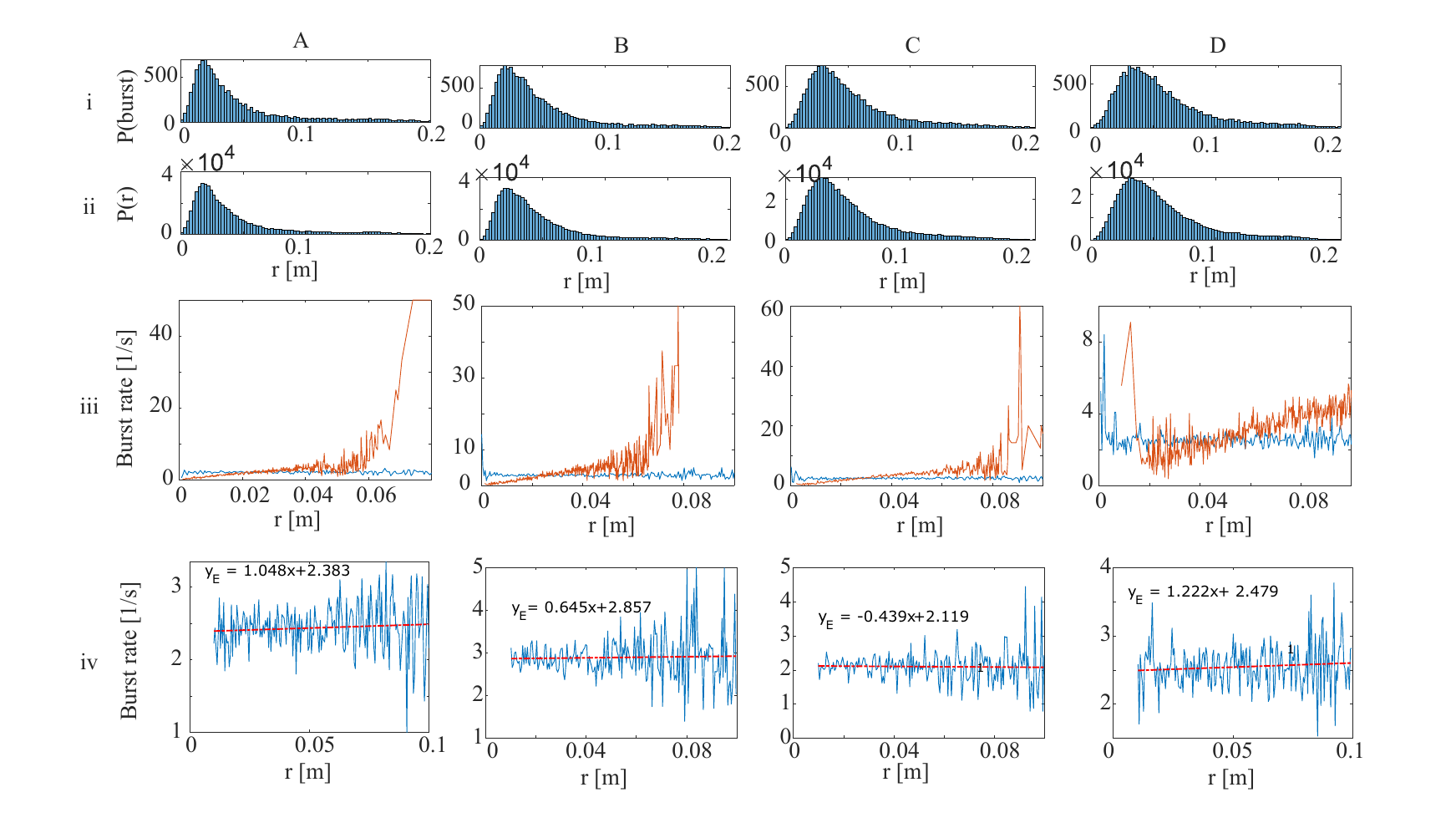}
    \caption{Rate to give a burst as function of the distance to the VF. (A) for $V_{VF}=0.04 [m/s]$, (B) for $V_{VF}=0.05 [m/s]$, (C) for $V_{VF}=0.06 [m/s]$, and (D) for $V_{VF}=0.07 [m/s]$. First row of plots show the probability to be at each distance, $P(r)$, and the probability to burst at each distance, $P(burst)$ (for the experimental data). For calculating the $P(burst)$ we collect distances where we have a burst, then bin the distances to $100$ bins, and for each of them calculate how many bursts occurred. Then we calculate $P(r)$ - the probability that the RF spend time in each of our bins, and we excluded from it the positions during the bursts (the time between burst and next peak of speed). The middle row shows the calculated burst rate, using Eq.\ref{fburst}, in experiments (blue) and the model simulations (orange). The last row of plots present a linear fit to the experimental rate to burst vs the  distance to the target.}
    \label{linear_kiind_of}
\end{figure}

In order to examine if the experimental tail bursts are more frequent as the distance to the target fish is larger, as assumed in our model (Eq.9), we collected all the bursts events (the minimums of the $V_{RF}$) and binned them as function of the distances ($|\vec{r}|$) in which they occur (Fig.\ref{linear_kiind_of}, $P(burst)$). The rate to give a burst is calculated as follows: we first computed the probability to be at each location (by excluding all the locations just after the bursts and until the maximum of the velocity), and then divided the probability to burst at each distance ($P(burst)$) by the probability to be at each distance ($P(r)$), and get the rate of the tail to burst at different distances: 
\begin{equation*}
    f_{burst} [\frac{1}{sec}]= \frac{P(burst)}{P(r)\Delta t }
    \label{fburst}
\end{equation*}
where $\Delta t=0.01 [s]$ is the time step and is the same in both the experiments and the simulations. To check whether the frequency is indeed higher at greater distances, we plot the rate ($f_{burst}$) for different distances, $|\vec{r}|$ (Fig.\ref{linear_kiind_of}), and we concluded that the frequency to exert a burst is higher for greater distances. However, the statistics from the experiments are pretty low, due to the short duration of the experimental trajectories. Also, the probability to be at edge positions, very far from the target (above $0.07[m]$) or very close to the target (below $0.02[m]$), is low, and still, we may get a burst event in those positions, resulting in large fluctuations in the calculated burst rate (dividing by a small number in Eq.\ref{fburst}). For the fastest VF, the linear relation is clear in the experimental data, while it is not very clear for the slower moving VF.

\section{Velocity threshold }\label{Vtreshold}

\begin{figure} [H]
    \centering
    \includegraphics{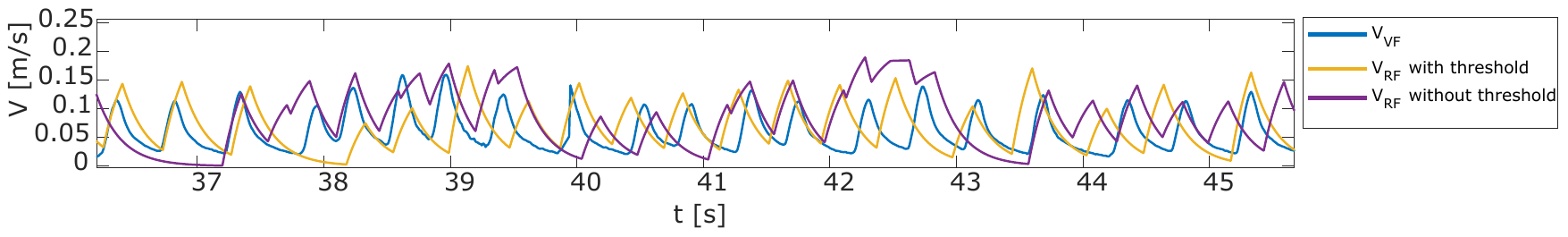}
    \caption{The RF velocity: the simulated RF's velocity dynamics without the velocity threshold is shown in purple, while the RF's velocity with the threshold ($0.04 [m/s]$) is in yellow. The experimental VF velocity is in blue. We used here the same parameters as in Fig.2 (average $V_{VF}=0.06[m/s]$).}
    \label{Odd}
\end{figure}

In order for the simulated tail bursts to occur in a realistic manner, we found it crucial to define the velocity threshold. Only when the RF's velocity is below the value of the threshold ($0.04 [m/s]$), a tail burst could occur (meaning we randomize a number - the tail burst event, as described in the previous section).  An example of the tail waiving model without such a threshold is presented in Fig.\ref{Odd}, in which it is clear that the tail bursts could happen too quickly from the previous peak without this threshold (see the purple line). 

\begin{figure} [H]
    \centering
    \includegraphics{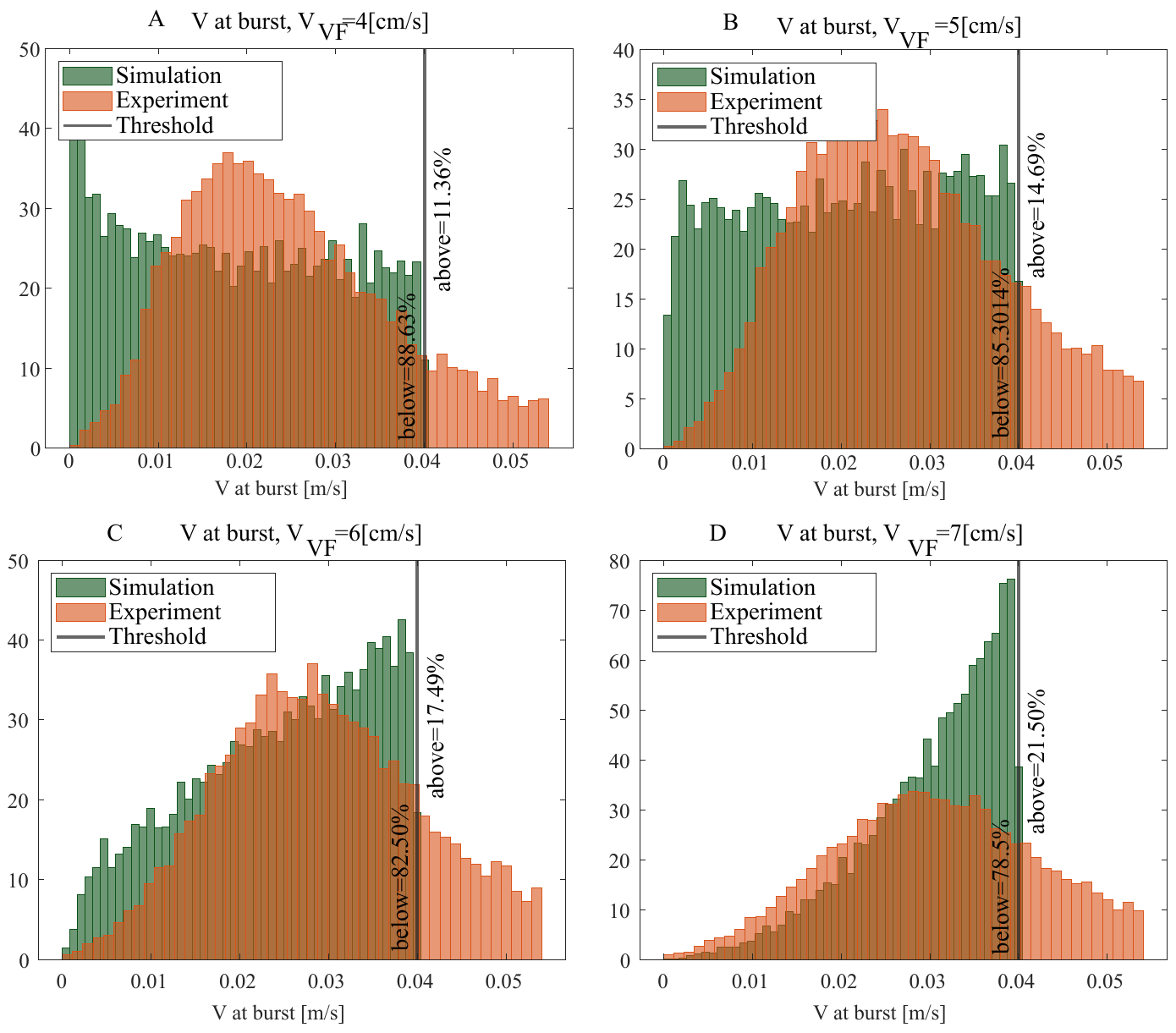}
    \caption{The distributions of the velocity at the onset of bursts for different VF velocities. The green distributions are for simulations and the orange distributions for the experiments. The black line is the velocity threshold we chose ($0.04 [m/s]$), below which most of the experimental bursts occur. }
    \label{treshold_example}
\end{figure}

We demonstrated why we chose a velocity threshold of $0.04 [m/s]$ in Fig.\ref{treshold_example}, where are shown  histograms of the experimental RF speed values at the burst events. The values of minima of the speeds (where the burst events occur) are mostly below our chosen threshold. When we tried a higher value the RF velocity dynamics didn't match the experiments.

\section{Distribution of the amplitude of the velocity bursts from the experimental RF dynamics}

We extracted the difference in the experimental RF velocity between the time of burst initiation and its maximum. 

\begin{figure}[H]
\centering
\includegraphics[]{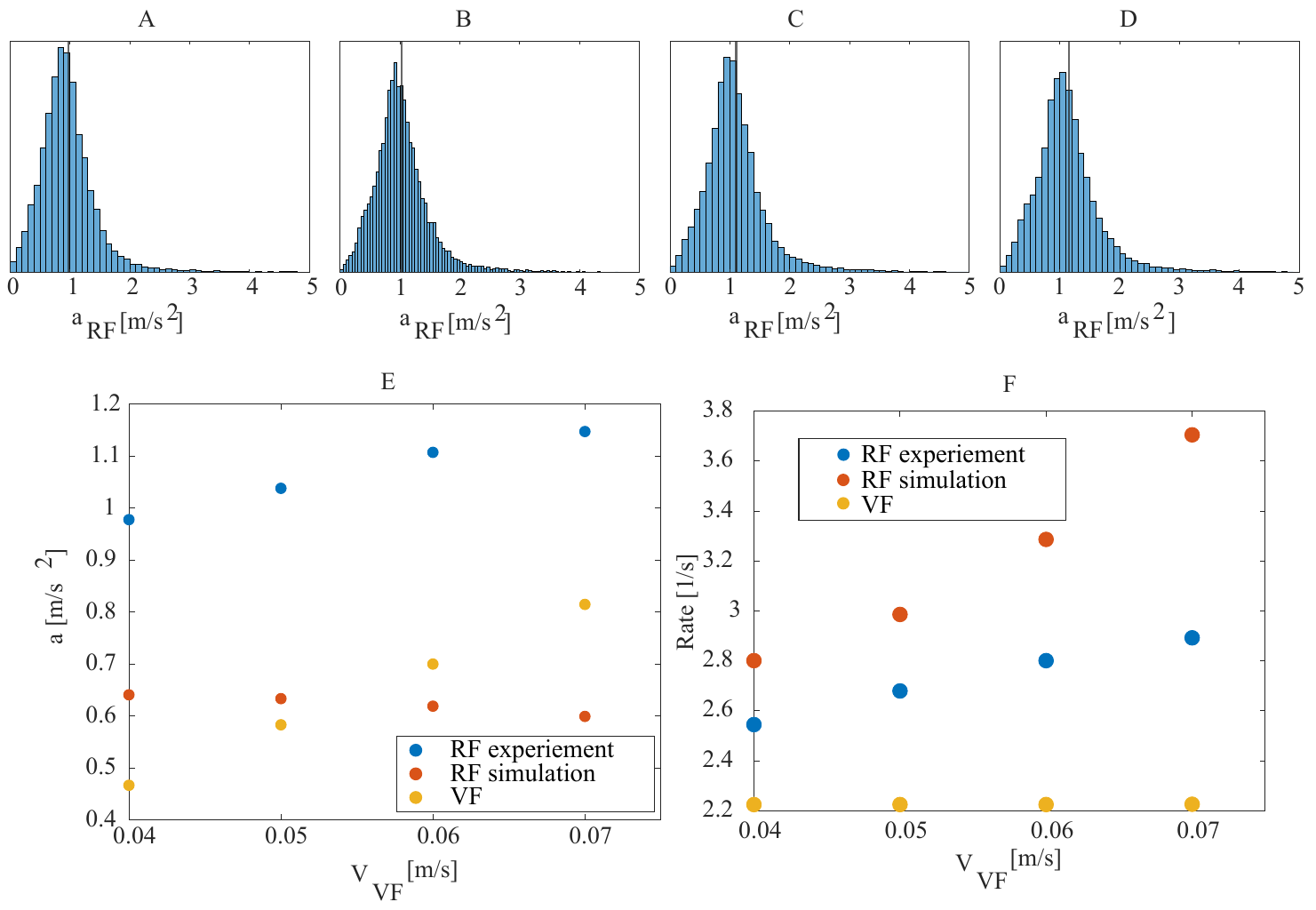}
\caption{Distributions of the RF's acceleration (burst amplitude divided by the time of each burst) from the experimental data, following 1VF: (A) $V_{VF}=0.04[m/s]$, (B) $V_{VF}=0.05[m/s]$, (C) $V_{VF}=0.06[m/s]$, and (D) $V_{VF}=0.07[m/s]$. (E) The average acceleration amplitude for the experimental RF (blue), simulated RF (red), and the VF (yellow). (F) The average rate of the tail bursts for the RF in experiments (blue), the simulated RF (red), and the VF (yellow). This rate is calculated as the average time duration between a burst event and the timing of the prior peak velocity.}
  \label{rates}
\end{figure}

\section{Attention threshold of the RF}\label{Attention}

By expanding the model from 1VF to several VF, we defined a threshold ($\tau$) of neural firing to determine which targets contribute to the calculation of the distance to targets, Eq.9. On each iteration, the neural firing ($n_i$) is monitored to determine if it is above the threshold. If so, the distance to the relevant VF is included in the calculation of $|\vec{r}|$ as follows: 

\begin{equation}
    |r|=\frac{\sum_{i=1}^{m} |\vec{r}_i|}{m}, \quad \text{for all}\ \frac{n_i}{N/k}> \tau
\label{weighted_r}
\end{equation}

where $m$ equals the number of neural groups which fire above the threshold. The $N$ is the total number of neurons in all the groups, $k$ is the number of neural groups, and $|\vec{r}_i|$ is the calculated distance of the RF to target $i$.

\begin{figure}[H]
\centering
\includegraphics[scale=.6]{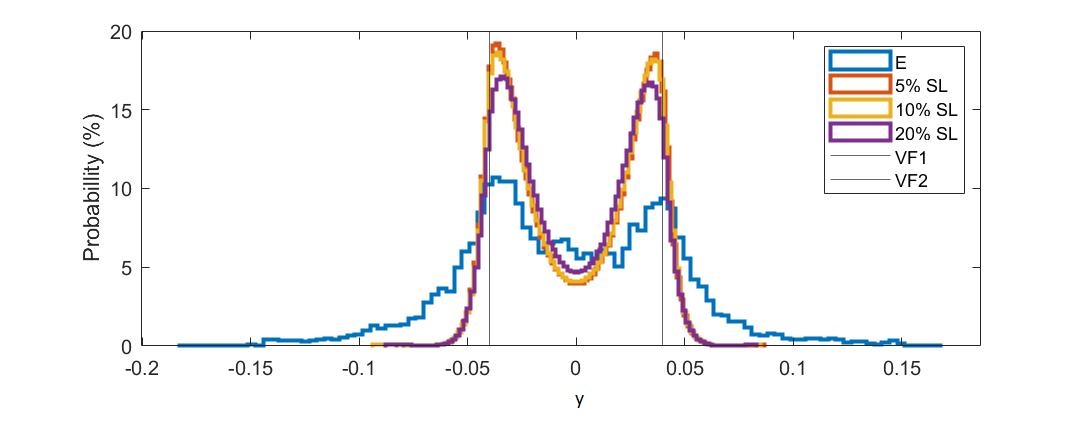}
\caption{\footnotesize{Examining the effect of different $\tau$ parameters on the probability distrubutions of $y_{RF} [m]$, the simulations ran for $V_{VF}=0.04 [m/s]$. Using the following parameters: $\beta=1 [1/s]$, $\gamma=5 [1/s]$, $\sigma=\pi [rad/s]$, $\xi=0.01 [m/s^2]$, $b=\pi [rad/s]$, $\nu=0.5$,  $r_d=0.2 [m]$, $\sigma_\theta=0.00002\pi$, and $k=0.95 [1/s]$.}}
\label{SI_attention_fig}
\end{figure}

In the case of the 2VF and 3VF, we included: $\tau=10\%$. As shown in Fig.\ref{SI_attention_fig} the RF's $y$-position behind the targets are not very sensitive to the threshold value.

\section{RF following 1VF}

\begin{figure} [H]
\centering
\includegraphics[]{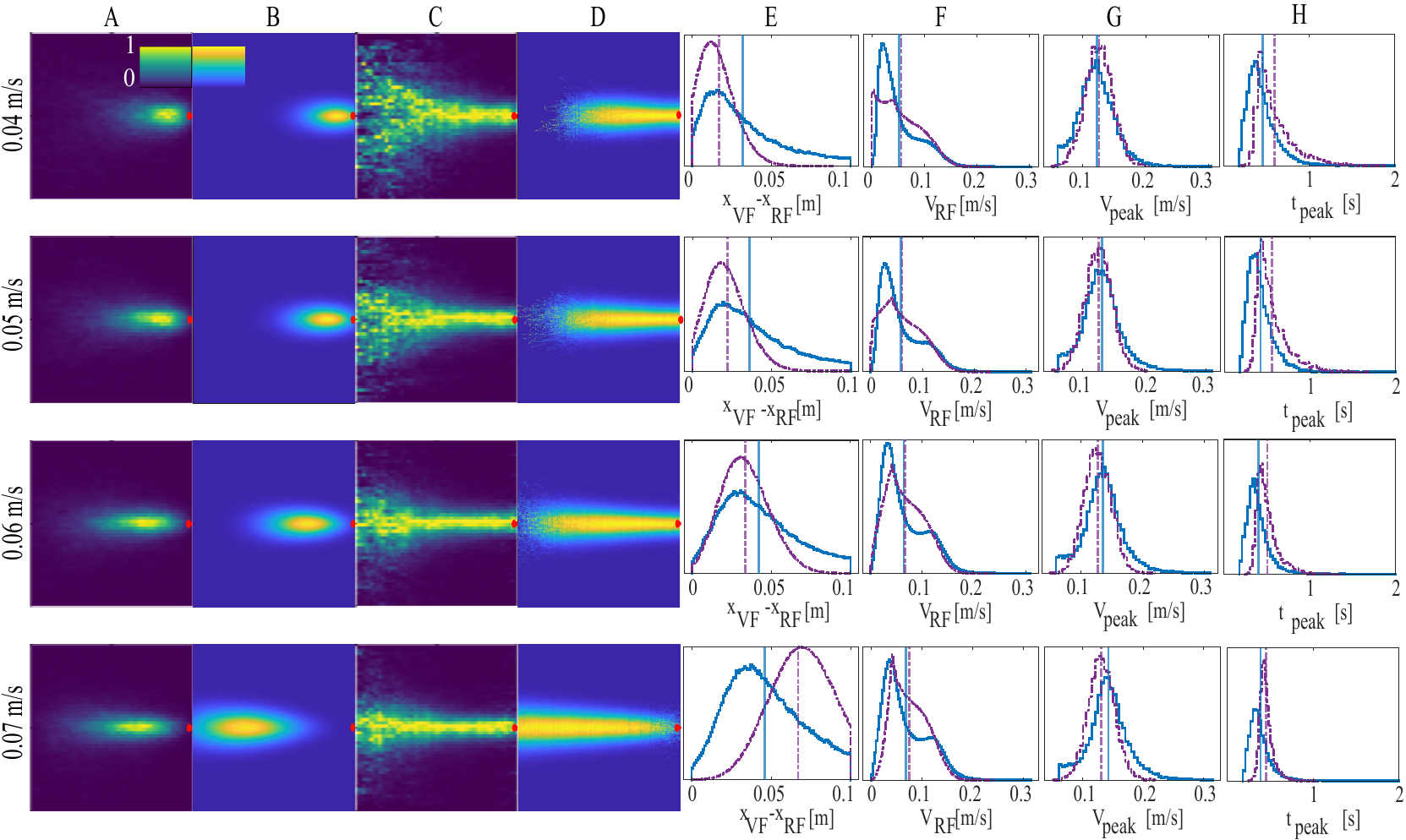}
\caption{RF chasing 1VF. (A-D) Accumulated distribution of the RF behind the VF, for different mean $V_{VF}$ (each row), in the VF frame of reference (the VF is at the origin, red dot). For each velocity we normalized the heat maps over the whole 2D space (A and B), or over individual lines of constant $x$-sections (C and D). (A) and (C) Displays the experimental heat maps, (B) and (D) shows the simulated results. (E-H) show the distributions (purple for simulation and blue for the experimental data) of the RF relative position along the $x$-axis (E), velocity (F), peak velocity values (G) and the consecutive time between velocity peaks (H). We used the same parameters as in Fig.2. For each VF velocity, we ran $100$ simulations (in which the RF initial position was random) each for $5000 [s]$ (500,000 iteration steps).}
\label{oneVF}
\end{figure}

We present here extended data for the case of the RF following 1VF (complimentary to Fig.2). For each velocity we compare the distribution of the distance between the RF and the VF along the $x$-axis (direction of motion of the VF, Fig.\ref{oneVF}E), which emphasizes that the RF lags further behind the VF as the VF speed increases. In our model this trend arises, as the tail-bursts occur more rapidly as the distance to the VF increases (Eq.9), enabling the RF to chase faster VF at larger lag distances.
At the fastest VF speed the model gives a lag distance that is significantly larger than the experiment. However, all other measures of the RF dynamics, as shown in Fig.\ref{oneVF}F-H, are in reasonable agreement across all VF speeds. Although a more precise fit could be obtained by adjusting the mean burst force $f_0$ parameter to each $V_{VF}$, we prefer to keep all the model parameters as constant, such that the model is as general as possible.

\section{RF following 2VF}
We present here extended data for the case of the RF following 2VF (complimentary to Fig.3).

Note that we could postulate that the fish makes weaker burst forces ($f_0$) when the frequency of its tail bursts decreases, which could improve the agreement, but we opted to keep the model as simple as possible, and maintain that all the parameters are independent of the average VF speed. In particular, we chose a burst force amplitude that would prevent the RF from losing the fastest moving VF ($0.07 [m/s]$), which causes it to move too close to the VF for the slower moving VF cases. 

\begin{figure}[H]
\centering
\includegraphics[]{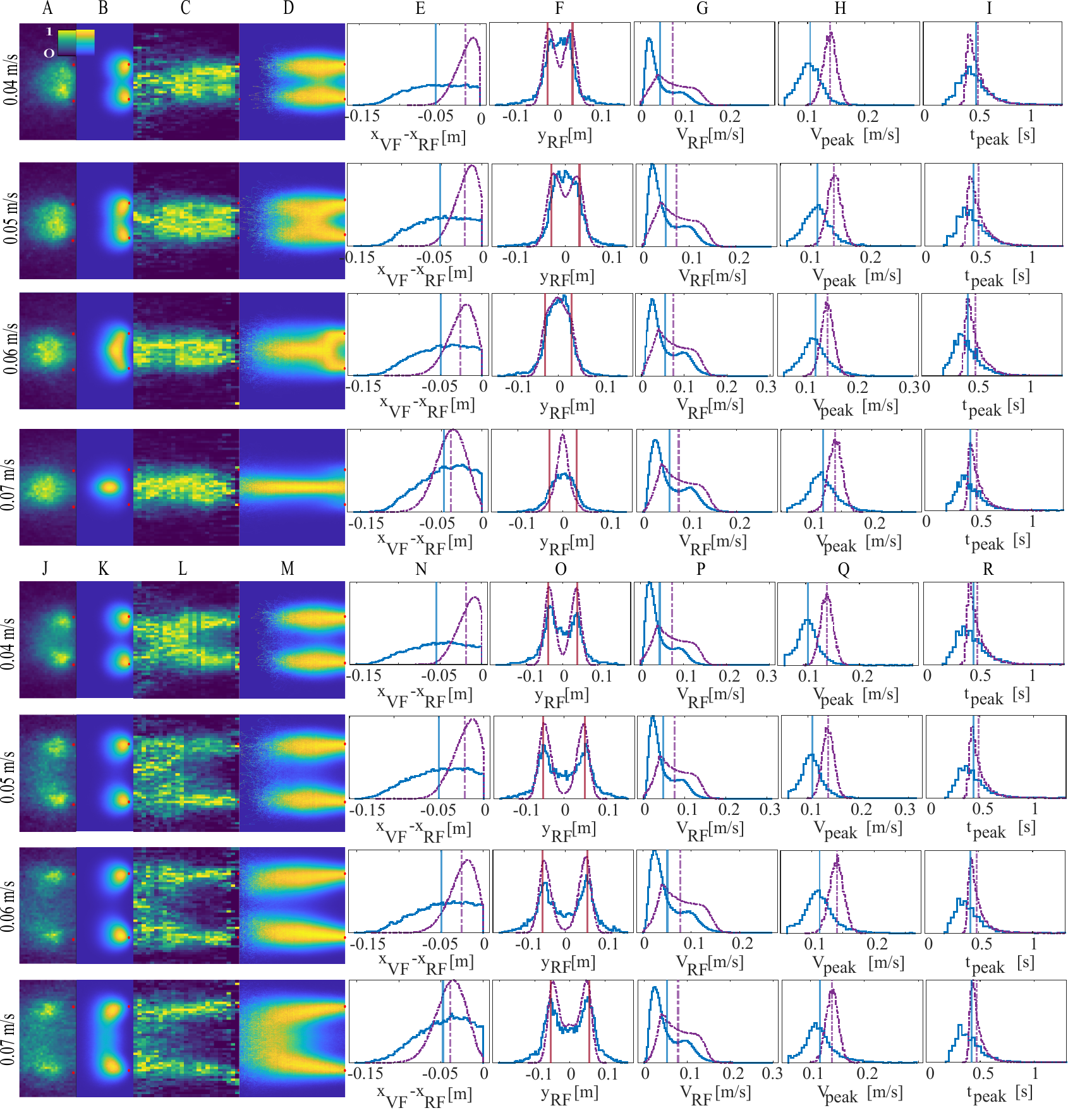}
\caption{Two VF - heat maps normalized over the whole 2D space (A-B and J-K) or over the $x$-axis sections (C-D and L-M). (A,J) and (C, L) The experimental RF's accumulated distribution, while (B, K) and (D, M) show the corresponding simulated distributions. (E, N) Distributions of the RF's $x$-positions relative to the 2VF, and $y$-positions (F, O). (G, P) Histograms of the RF speed ($V_{RF}$), (H) burst velocity peak values ($P_{peak}$), and (I) the time intervals between consecutive velocity peaks ($t_{peak}$). All panels compare the experimental data (blue) with the model (purple). The average of each distribution is given by the vertical line, with the corresponding color. The top four lines are for $LRD=0.06 m$, while the bottom four lines are for: $LRD=0.08[m]$ for $V_{VF}=0.04[m/s]$, $LRD=0.1[m]$ for $V_{VF}=0.05[m/s]$, and $LRD=0.11[m]$ for $V_{VF}=0.06[m/s],0.07[m/s]$). We used the same parameters as in the 1VF system (Fig.2), except for: $f_0=1.2[m/s^2]$. When dealing with more than one VF we also use the following parameters: $\nu=0.5$ and $\sigma_\theta=0.00002\pi$. For each VF velocity, we ran $100$ simulations (in which the RF initial position was random) each for $5000 [s]$ (500,000 iteration steps).}
\label{twoVF}
\end{figure}

\section{RF following 3VF}
We present here extended data for the case of the RF following 3VF (complimentary to Fig.4).

\begin{figure}[H]
\centering
\includegraphics[]{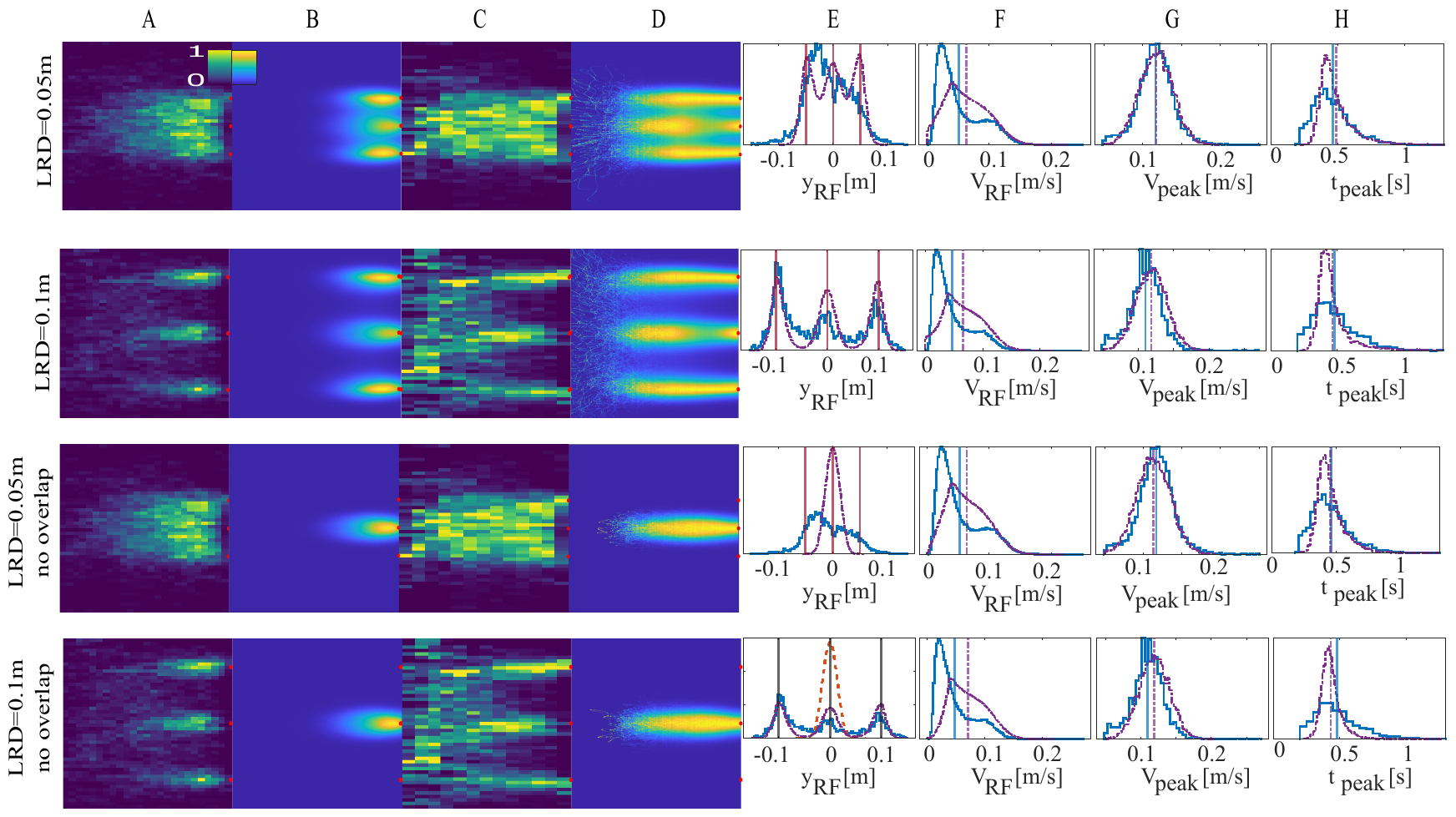}
\caption{RF following 3VF. The LRD value for each line in (A-H) is given on the left, with the lowest two lines presenting simulations without the overlap function. Accumulated spatial RF distribution, normalized over the whole 2D space (A,B) or over the $x$-axis sections (C,D). (A,C) Experimental data, while (B,D) the simulation results. (E) Distributions of the RF's projected $y$-positions, (F) RF's speed, (G) Speed of the RF at the peaks, and (H) the time intervals between consecutive speed peaks. In (E-H) we compare the experimental data (blue) with the model (purple). The average of each distribution is denoted by the vertical line, with the corresponding color. In (A-H) the VF velocity is $0.04 [m/s]$. We used the same parameters as in Fig.2, except for the burst force amplitude which was changed to $f_0=0.95 [m/s^2]$. we ran $100$ simulations (in which the RF initial position was random) each for $5000 [s]$ (500,000 iteration steps).}
  \label{3VF}
\end{figure}

The distributions of the RF speed are in reasonable agreement between the simulations and the experiments (Fig.\ref{3VF}F), while the distributions of the velocity peaks and the time intervals between consecutive velocity peaks are in very good agreement (Fig.\ref{3VF}G,H).

\section{RF following 2VF in shifted geometry}
We present here extended data for the case of the RF following 2VF in shifted geometry (complimentary to Fig.5).

\begin{figure}[H]
\centering
\includegraphics[]{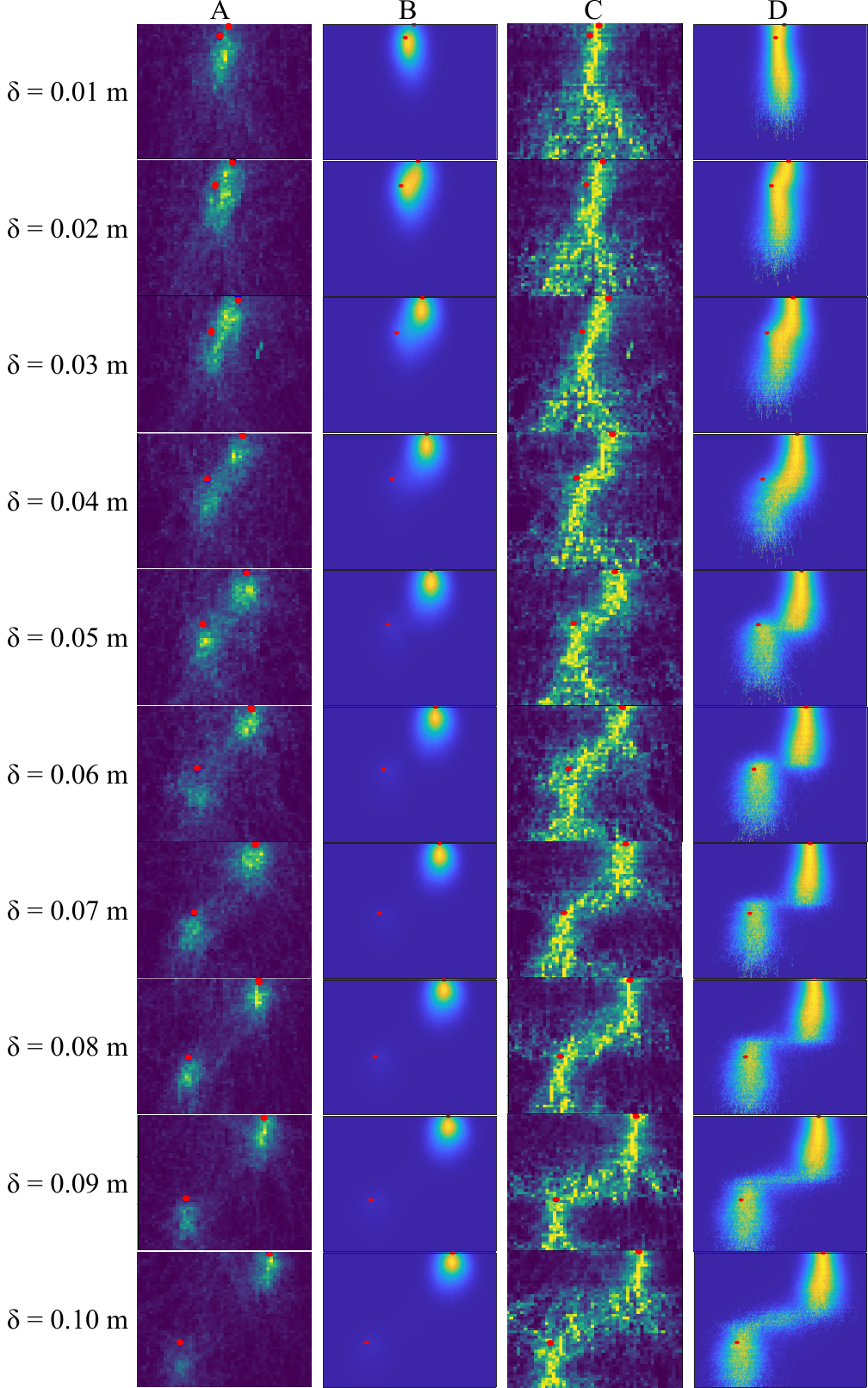}
\caption{2VF in a shifted geometry: one of the VF is shifted in the front-back and the left-right directions with respect to the other VF, by $\delta$. The accumulated spatial distribution of the RF behind the 2VF, (A,B) normalized over the whole 2D space, or (C,D) normalized over $x$-axis sections. (A) and (C) show the results from experimental data, while (B) and (D) show results from the simulations. The VF are denoted by the red circles. The model parameters are the same as used in Fig.\ref{twoVF}. The right panels show examples of RF trajectories relative to the 2VF (red dots), for the cases of $\delta=0.03 [m]$, and $\delta=0.06 [m]$. In both plots the color of the line represent the time $[s]$.}
  \label{Symy}
\end{figure}

\section{Many RF}

\begin{figure} [H]
    \includegraphics{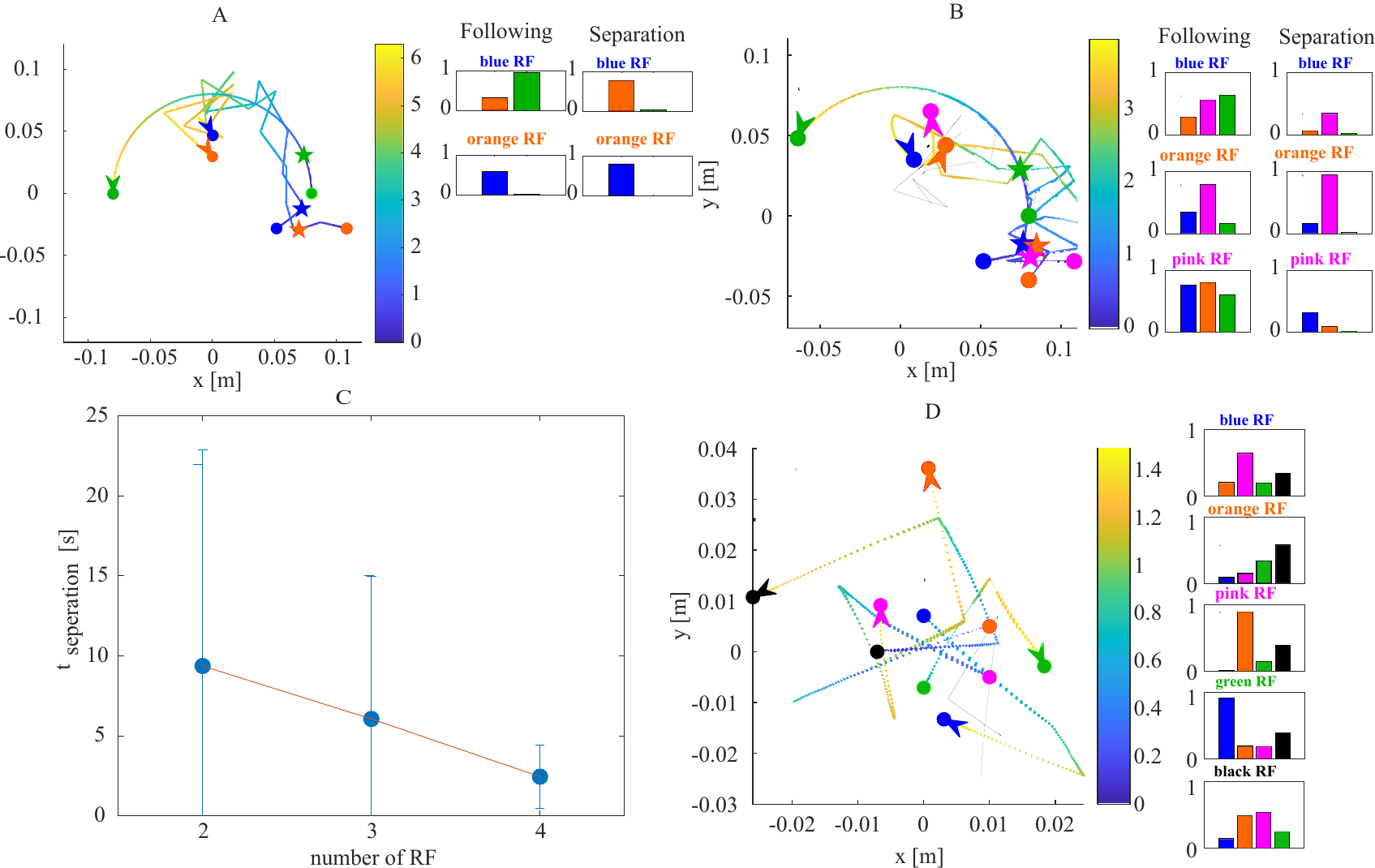}
    \caption{Simulating groups of RF. (A, B) Examples of simulated trajectories, with (A) 2RF and (B) 3RF in the presence of a VF (green) swimming in a circle of radius $0.08 [m]$ with a constant velocity of $0.05[m/s]$. The starting positions of the RF and VF are denoted by the colored circles, and the end points by the colored fish. The side panels show the spin states of each of the RF, representing the weight of the fellow RF and the VF on their direction of motion. We denote these spin states at a typical "following state" (denoted by the stars on the trajectories) where the RF were following the VF, and the "separation state" at the end of the trajectories, where the RF formed a shoal and stopped following the VF. (C) A plot of the mean time until the RF separate from the VF, as a function of the number of RF, based upon running $90,000$ simulations for each case of RF number. The mean times for 2RF,3RF and 4RF  are: $t(2RF)=9.36[s]\pm 13.49[s]$, $t(3RF)=6.05[s]\pm 8.89 [s]$ and $t(4RF)=2.44[s]\pm 1.96 [s]$. (D) A typical short trajectory of five RF swimming together with no leader (colored circles denote the starting points). We used the same parameters as in Fig.2.}
    \label{separation}
\end{figure}

\begin{figure} [H]
    \centering
    \includegraphics{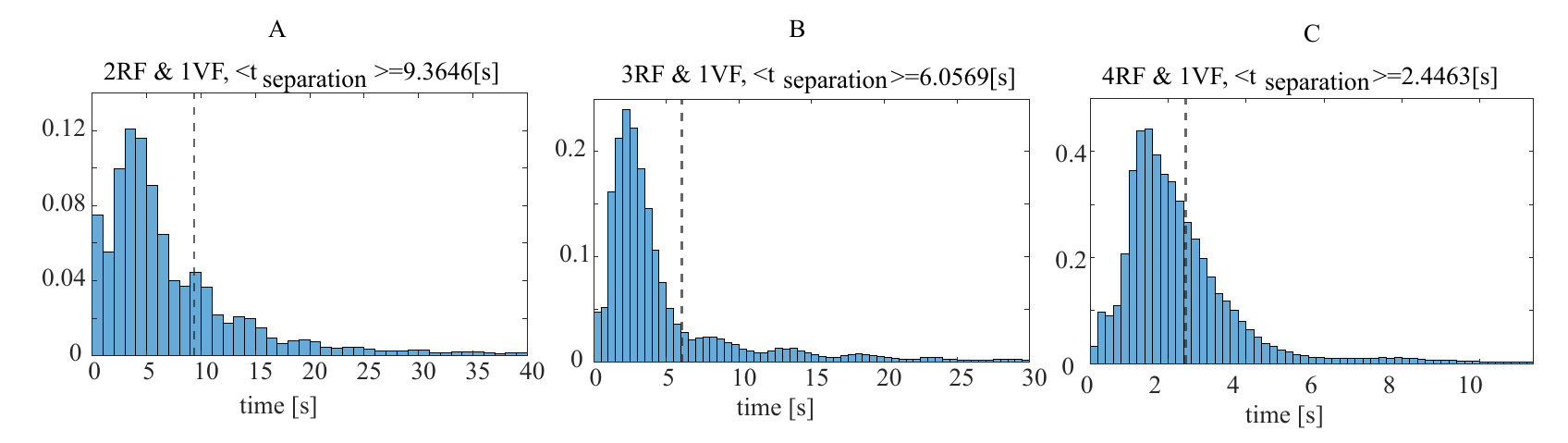}
    \caption{Separation time histograms for many RF and one VF - the plots show the time of separation for 2RF (A), 3RF (B) and 4RF (C) that are following one VF in a circle of radius $0.08 [m]$ with a linear velocity $V_{VF}=0.05 [m/s]$ (Fig.6A,B). The average separation times (Fig.6C) were calculated from conducting $90,000$ simulations. All simulations ran with the same parameters as in the 1VF case (Fig.2). }
    \label{many}
\end{figure}

We may now ask what happens when a group of RF interact with each other according to our model. In Fig.\ref{separation}A,B we plot examples of typical trajectories for either 2RF or 3RF, in the presence of a single persistent leader, in the form of a single VF that moves along a circle (green, see also Supplementary Movies M7-M9). The RF start at the positions indicated by the full circles. We find that at an early time the RF follow the VF, as indicated by the "Following" state (at the locations indicated by the stars). At the corresponding times the histograms on the right show the proportion of "on" spins in each RF, indicating that the blue RF has its maximal attention on the VF leader, while the orange RF is following the blue RF. This continues for a certain time, until their attention changes to following each other. This occurs at the time indicated by the fish shapes on the trajectory, and the corresponding histograms shown at this "separation" time. The RF lose the leader, and form a non-polar shoal. The time duration until this separation event occurs decreases with increasing number of RF, as shown in Fig.\ref{separation}C (the full distributions of the separation times are given in Fig.\ref{many}).

In Fig.\ref{separation}D we demonstrate that in the absence of any leader fish, our model naturally gives rise to cohesive shoal behavior, without any spontaneous alignment of the fish along a particular direction (see Supplementary Movie M10). This falls into the category of "swarm" behavior, with low global polarization of the velocity vectors \cite{delcourt2012shoals}. Note that our model ignores the finite size of the RF, which means that we do not describe the short-range maneuvers that are necessary for collision avoidance. Similarly, we ignore obstructions of one RF by another, with all the RF visible at all times. In addition, our model explicitly involves many-body interactions, that are not simply the superposition of pairwise interactions. Significant interactions beyond pairwise have indeed been measured inside fish schools \cite{katz2011inferring}.

\section{Supplementary Movies}
The simulated RF dynamics while following 1VF, 2VF, and 3VF are shown in movies M1-M6. The many RF following 1VF target moving along a circle are shown in movies M7 (2RF), M8 (3RF), and M9 (4RF). Also, we show the many RF dynamics with no leader among them in movie M10 (5RF). All movies are in the following link: 
\href{https://drive.google.com/drive/folders/1H76yOawtv9zuou84VY2HrxOjO-2jXu-w?usp=sharing}{Movies} 
\begin{itemize}

\item Movie M1- one RF following one VF ($V_{VF}=0.06[\frac{m}{s}]$), the top plot shows the RF in the real frame of reference, and the bottom plot, the RF is simulated relative to the VF (which is in the origin). The RF is in blue and the VF is in red in both cases (parameters as in Fig.2).

\item Movie M2 - one RF (blue) following 2VF (red) in the compromise regime ($LRD=0.06[m]$, and $V_{VF}=0.07[\frac{m}{s}]$), the right plot shows the neuronal firing in the RF brain, in which "1" represent neurons directed at the VF in $y=0.03 [m]$, and "2" represent neurons directed at the VF in $y=-0.03 [m]$ (parameters as in Fig.3).

\item Movie M3- one RF (blue) following 2VF (red) in the bifurcated regime ($LRD=0.11[m]$, and $V_{VF}=0.07[\frac{m}{s}]$), the right plot shows the neuronal firing in the RF brain, in which "1" represent neurons directed at the VF in $y=0.055[m]$, and "2" represent neurons directed at the VF in $y=-0.055[m]$ (parameters as in Fig.3).

\item Movie M4- one RF (blue) following 2VF (red) in the shifted configuration ($LRD=0.03[m]$, $V_{VF}=0.04[\frac{m}{s}]$, and $\delta=0.03[m]$), the right plot shows the neuronal firing in the RF brain, in which "1" represent neurons directed at the VF in $y=0.03[m]$, and "2" represent neurons directed at the VF in $y=-0.03[m]$ (parameters as in Fig.3).

\item Movie M5- one RF (blue) following 3VF (red) in the compromise regime ($LRD=0.05[m]$, and $V_{VF}=0.04[\frac{m}{s}]$), the right plot shows the neuronal firing in the RF brain, in which "1" represent neurons directed at the VF in $y=0.05[m]$, "2" represent neurons directed at the VF in $y=0[m]$, and "3" represent neurons directed at the VF in $y=-0.05[m]$ (parameters as in Fig.4).

\item Movie M6- one RF (blue) following 3VF (red) in the bifurcated regime ($LRD=0.10[m]$, and $V_{VF}=0.04[\frac{m}{s}]$), the plot on the right side shows the neuronal firing in the RF brain, in which "1" represent neurons directed at the VF in $y=0.1 [m]$, "2" represent neurons directed at the VF in $y=0[m]$, and "3" represent neurons directed at the VF in $y=-0.1[m]$ (parameters as in Fig.4).

\item Movie M7- 2RF following 1VF (green) in a circle (of radius $0.08[m]$). The neuronal activity for each RF are given in the right bar plots, where the bar color indicates a neuronal group directed at the fish with the same color (parameters as in Fig.2). 

\item Movie M8- 3RF following 1VF (green) in a circle (of radius $0.08[m]$). The neuronal activity for each RF are given in the right bar plots, where the bar color indicates a neuronal group directed at the fish with the same color (parameters as in Fig.2). 

\item Movie M9- 4RF following 1VF (green) in a circle (of radius $0.08[m]$). The neuronal activity for each RF are given in the right bar plots, where the bar color indicates a neuronal group directed at the fish with the same color (parameters as in Fig.2). 

\item Movie M10- Simulation of 5RF (a shoal) with no leader. The right plot shows the polarization vector- the average direction of the whole shoal (parameters as in Fig.2).
\end{itemize}

\bibliographystyle{apsrev4-1}
\bibliography{SI}